\documentclass[a4paper,11pt]{article}
\usepackage{jheppub} 
\usepackage{amssymb}
\usepackage{booktabs}
\usepackage{float}

\newcommand{\KK}{\mathrm{KK}}
\newcommand{\MPl}{M_{\rm Pl}}
\newcommand{\Tr}{\mathrm{Tr}}
\newcommand{\dd}{\mathrm{d}}
\makeatletter
\def\@fpheader{\relax}
\makeatother

\title{Momentum-space entanglement and mutual information in Kaluza–Klein reduction}

\author[1]{Udaykrishna Thattarampilly\note{Corresponding author.}}
\author{and Yunlong Zheng}
\affiliation{Center for Gravitation and Cosmology, College of Physical Science and Technology,\\
Yangzhou University, Yangzhou 225009, China}

\emailAdd{uday7adat@gmail.com}
\emailAdd{zhyunl@yzu.edu.cn}

\abstract{In theories with compact extra dimensions, the four-dimensional effective description is obtained by Kaluza--Klein reduction. 
The state available to a four-dimensional observer is the reduced density matrix obtained by tracing the higher-dimensional vacuum over the tower of massive modes, and it need not be pure. In this paper we study the momentum-space entanglement of this state for a five-dimensional scalar on $M_4\times S^1$ with a cubic or quartic self-interaction. We compute, at leading order in the coupling, the mutual information $I_n$ between the zero modes below a resolution scale $\mu$ and a single Kaluza--Klein level $n$. When the zero modes couple linearly and without derivatives to pairs of tower modes, $I_n$ falls as $1/m_n$, where $m_n$ is the mass of level $n$. In the cubic theory this tail is independent of the ultraviolet cutoff, with coefficient $(\mu R)^2/1024\pi^3$ for compactification radius $R$, and the mutual information with the whole tower grows logarithmically with the number of levels. In the quartic theory the tail is masked by states containing additional hard quanta, which make $I_n$ depend on the cutoff. Compared with the corrections the tower makes to low-energy amplitudes, which are dominated by the lightest levels and captured by local operators, the mutual information with the tower is not dominated by any finite set of levels and arises from the time-nonlocal part of the effective action. The local four-dimensional effective theory therefore describes the low-energy observables of the reduced system but not its state. Unbroken supersymmetry cancels the vacuum energy of the compactification but not the mutual information, which at leading order is a sum of squared amplitudes.}

\begin{document}
\maketitle
\flushbottom

\section{Introduction}
\label{sec:introduction}

Theories with extra dimensions are a standard part of high-energy physics. They arise in string theory and supergravity, and they underlie many phenomenological models of physics beyond the Standard Model \cite{Appelquist:2000nn}, including solutions of the hierarchy problem \cite{Arkani-Hamed:1998jmv,Randall:1999ee}. The low-energy content of such theories is obtained by Kaluza--Klein (KK) reduction \cite{Kaluza:1921tu,Klein:1926tv}. The spacetime is of the form $M_4\times K$, where $K$ is a compact internal manifold, and each field is expanded in eigenmodes of $K$. The zero modes are the light fields accessible to a four-dimensional observer at energies below the compactification scale, while the massive modes form a tower whose masses are determined by the size of $K$. The tower is then integrated out, and the reduction is justified by the decoupling of the tower from the zero modes. By the Appelquist--Carazzone theorem \cite{Appelquist:1974tg}, states of mass $m_n$ affect low-energy observables  only through local operators suppressed by powers of $1/m_n$. At energies well below the compactification scale the tower can therefore be replaced by a local four-dimensional effective Lagrangian \cite{Georgi:1993mps,Manohar:2018aog}.

Concepts from quantum information have meanwhile become standard tools in quantum field theory \cite{Casini:2009sr,Nishioka:2018khk,Eisert:2008ur}. Most of this work concerns entanglement between regions of space. A low-energy observer, however, is limited by momentum resolution rather than by access to a region of space. This observer has access to modes below a certain momentum scale, and his description of the system is given by the reduced state obtained by tracing out the rest. 

Momentum-space entanglement was studied systematically by Balasubramanian, McDermott and Van Raamsdonk (BMV) \cite{Balasubramanian:2011wt}. They computed, in perturbation theory, the entanglement entropy and mutual information between modes at different momentum scales, and found that the entanglement decreases as the scales separate, suggesting that this decay is related to decoupling. They also pointed out that the reduced density matrix of the low-momentum modes is naturally associated with the the Wilsonian effective action \cite{Wilson:1973jj,Polchinski:1983gv}. This action is in general nonlocal in time, and its nonlocal terms are necessary for the entanglement \cite{Costa:2022bvs,MartinsCosta:2022bjr}, so the local effective action obtained by a derivative expansion leaves the reduced state pure. Momentum-space entanglement has since been studied for scalar fields on fuzzy spheres \cite{Kawamoto:2021rox} and for interacting fermions at finite density \cite{Hsu:2012gk}. The purity of the reduced state has been used to diagnose the breakdown of perturbation theory and to derive unitarity bounds \cite{DuasoPueyo:2024usw}. 

Momentum-space entanglement has not yet been studied in the context of dimensional reduction. In this scenario the four-dimensional observer has access to the zero modes but not to the tower, and the corresponding state is the reduced density matrix obtained by tracing the higher-dimensional vacuum over the tower. Decoupling guarantees that the observables computed from this state are reproduced by a local four-dimensional theory, but it does not require the state itself to be pure.\footnote{Balasubramanian, McDermott and Van Raamsdonk call entanglement entropies and mutual informations natural observables of quantum field theory~\cite{Balasubramanian:2011wt}. In this paper we use the word in the narrower sense of low-energy observables, such as correlation functions at energies well below the compactification scale. Entropies and mutual informations are nonlinear functionals of the state and are not of this form (Section~\ref{sec:local}).} This partition differs from those studied previously. The reduction discards entire KK levels, and each level is a four-dimensional field of mass $m_n=|n|/R$ with a wave mode corresponding to every three-momentum. Discarding a level therefore removes all of these modes at once, rather than a single mode or a shell of momenta. The masses of the levels and their degeneracies are fixed by the geometry of the internal manifold rather than by an adjustable scale, and invariance under translations along the internal directions implies conservation of KK number, $\sum_in_i=0$ at every vertex, which restricts the processes that can entangle the zero modes with the tower. 


In this paper we compute the entanglement entropy of the zero-mode sector and its mutual information with the tower for a real scalar on $M_4\times S^1$ with a cubic or a quartic interaction, at leading order in the coupling. For free fields the reduction is lossless. Once interactions are switched on, however, the reduced state is mixed. The entropy $S_A$ of the accessible sector (zero modes with momenta below a resolution scale $\mu$) diverges at leading order in the quartic theory, as BMV showed for four-dimensional $\lambda\phi^4$. We therefore compute the mutual information $I_n\equiv I(A:B_n)$ between the accessible sector and KK level $\pm n$. It isolates the contribution of a single level and is insensitive to the purely zero-mode configurations. 

We find that, when the accessible sector couples linearly and without derivatives to pairs of tower modes, the mutual information $I_n$ with level $n$ falls as $1/m_n$. The contribution of a pair of modes of fixed momentum falls as $m_n^{-4}$, similar to the decline with scale separation found by BMV \cite{Balasubramanian:2011wt}. 
The mutual information with the tower therefore grows as $\ln N$ with the number of levels $N$. 

In the quartic theory, states with two soft zero-mode quanta and a tower pair have the same $1/m_n$ tail. The quartic vertex, however, also produces states in which the tower quanta are accompanied by a hard zero-mode quantum or by other tower modes. These states leave a second hard momentum free, so their contribution is dominated by momenta near the cutoff and outweighs the tail. As a result, $I_n$ depends on the cutoff. 

The corrections the tower makes to low-energy amplitudes fall as $1/m_n^2$, are dominated by the lightest levels, and are captured by local operators in the effective theory. Low-energy observables are therefore insensitive to how far the tower extends. The mutual information with the tower, in contrast, receives an equal contribution from every range of levels between $n$ and $2n$, so its total grows as $\ln N$ and depends on where the tower ends. This dependence is invisible to the effective theory. The entanglement arises from the terms of the effective action that are nonlocal in time, which a derivative expansion discards. It is measured by entropies and mutual informations, which are nonlinear functionals of the reduced density matrix. The local four-dimensional effective theory thus describes the observables of the reduced system but not its state and the missing information grows with the number of levels that are integrated out. Supersymmetry cancels the vacuum energy of the compactification, but at leading order it cannot cancel the entanglement, which is a sum of squared amplitudes. A supersymmetric compactification with vanishing vacuum energy therefore still leaves the four-dimensional observer in a mixed state, entangled with the tower at least as strongly as in the bosonic theory when its bosonic amplitudes are unchanged.

The couplings of the graviton and gauge zero modes to the tower have the same field content as the cubic vertex but carry derivatives of the tower fields. Power counting then places their mutual information with the tower dependent on the cutoff; a gauge-invariant calculation is needed to settle the question, and we leave it open.

The paper is organised as follows. Section~\ref{sec:background} introduces the setup and Section~\ref{sec:setup} obtains the spectrum of the reduced density matrix at leading order. Sections~\ref{sec:cha} evaluate the channels of the quartic theory, and Section~\ref{sec:mutual} defines the mutual information with a level and evaluates it in that theory. Section~\ref{sec:phi3} treats the cubic theory, including the sum over levels and the entropy of the accessible sector. Section~\ref{sec:evsi} compares the vacuum energy and the mutual information under supersymmetry, Section~\ref{sec:eft} discusses the implications for the four-dimensional description, and Sections~\ref{sec:discussion} and~\ref{sec:conclusions} contain the discussion and conclusions. Technical details are collected in the appendices.

\section{Background}
\label{sec:background}

This section is a brief review of existing results on KK reduction and momentum space entanglement entropy.

\subsection{Kaluza--Klein reduction}
\label{sec:kkprimer}

Consider a five-dimensional spacetime where the fifth dimension is compactified on a circle of radius $R$. A 5D real scalar field $\Phi(x,y)$ can be decomposed into a Fourier series along the compact coordinate $y \sim y + 2\pi R$:
\begin{equation}
\Phi(x,y)=\frac{1}{\sqrt{2\pi R}}\sum_{n\in\mathbb{Z}}\phi_n(x)\,e^{iny/R}, \qquad \phi_{-n}=\phi_n^\dagger .
\label{eq:modeexpansion}
\end{equation}
From the four-dimensional perspective, each Fourier coefficient $\phi_n(x)$ behaves as a scalar field with a mass $m_n = |n|/R$.
The $n=0$ mode is massless, while the $n \neq 0$ modes form an infinite tower of massive Kaluza--Klein (KK) levels. At energies $E \ll 1/R$, the heavy tower decouples \cite{Appelquist:1974tg}, and the low-energy physics is described by an effective 4D theory of the massless zero mode. This framework underpins numerous phenomenological models, from large extra dimensions \cite{Arkani-Hamed:1998jmv,Antoniadis:1998ig} to warped geometries \cite{Randall:1999ee}.

As a consequence of translation invariance along the compact direction, the KK label acts as a conserved charge, so every vertex must obey $\sum_i n_i = 0$.
KK excitations can be created only in combinations with vanishing total KK number. The level itself, however, is not a single mode. Fixing $n$ fixes only the mass of the corresponding four-dimensional field, while its three-momentum $\mathbf{q}$ remains continuous, with energy $\Omega = \sqrt{\mathbf{q}^2 + m_n^2}$. Tracing out a KK level therefore discards an entire relativistic species rather than a discrete set of states. 

\subsection{Reduced states, entropy and mutual information}
\label{sec:entprimer}

Let the full Hilbert space factorise as $\mathcal H=\mathcal H_A\otimes\mathcal H_B$, and suppose the global state $|\Omega\rangle$ is pure. An observer who has access only to the sector $A$ will perceive a state given by the reduced density matrix
\begin{equation}
\rho_A=\Tr_B\,|\Omega\rangle\langle\Omega| .
\end{equation}
For a pure global state, the von Neumann entropy
\begin{equation}
S_A=-\Tr\!\left(\rho_A\ln\rho_A\right)\ \ge\ 0
\end{equation}
 measures entanglement between $A$ and $B$. Alternative diagnostics include the purity $\Tr\rho_A^2\le1$ and the associated second R\'enyi entropy $S_2=-\ln\Tr\rho_A^2$. The purity equals one and $S_2$ vanishes exactly when $S_A$ does, but $S_2$ in general takes a different value. These are standard constructions; see \cite{Nielsen:2012yss} for the quantum-information background and \cite{Casini:2009sr,Nishioka:2018khk,Eisert:2008ur} for reviews aimed at field theorists.

In a field theory with several sectors, however, the total impurity of one sector is often less informative than its correlation with a specified part of the environment. For a tripartite factorisation $\mathcal H_A\otimes\mathcal H_B\otimes\mathcal H_C$, the mutual information
\begin{equation}
I(A:B)=S_A+S_B-S_{A\cup B}\ \ge\ 0
\end{equation}
measures the total classical and quantum correlation between $A$ and $B$. Unlike $S_A$, it does not count the correlations that either subsystem has with a third factor $C$. It also has better ultraviolet behaviour than the entropy of a single subsystem when the subsystems are chosen appropriately, a point illustrated in the momentum-space context by BMV \cite{Balasubramanian:2011wt}. This is the reason for the choice of mutual information as the quantity evaluated in this paper (Section~\ref{sec:mutual}).

These definitions do not require $A$ and $B$ to be spatial regions. Any tensor-factor decomposition of the Hilbert space defines a legitimate bipartition. A partition along momentum modes is the one relevant to Kaluza--Klein reduction, where the accessible zero-mode sector is separated from the heavy tower that is traced out.

\subsection{Momentum-space versus position-space partitions}
\label{sec:posspace}

The most familiar entanglement entropy is that of a spatial region. In the vacuum of a local quantum field theory it obeys an area law, $S\sim\mathcal A/\epsilon^2$, where $\mathcal A$ is the area of the region's boundary and $\epsilon$ is a short-distance cutoff \cite{Bombelli:1986rw,Srednicki:1993im,Casini:2009sr}. This behaviour is generic to the ground states of local Hamiltonians \cite{Hastings:2007iok,Eisert:2008ur}.

A Kaluza--Klein tower does contribute to this entropy, but its contribution says nothing specific about the reduction. The divergent area-law term is absorbed into a renormalisation of Newton's constant, through the mechanism of Sakharov and of Susskind and Uglum \cite{Sakharov:1967nyk,Susskind:1994sm,Visser:2002ew,Solodukhin:2011gn}. For a Kaluza--Klein tower, the resulting shift is exactly the species counting that relates the higher- and lower-dimensional Planck scales \cite{Dvali:2007hz,Dvali:2008jb,Solodukhin:2011gn}.

Position-space entanglement therefore measures mainly how many light species the spectrum contains. It does not measure how much correlation the four-dimensional observer loses by declaring those species unobservable. That loss depends on the state, and is captured by the momentum-space partition. 


\subsection{For free fields the reduction is lossless}
\label{sec:free}

It is useful to note that Kaluza--Klein reduction does not generate entanglement in a free theory. This establishes that any information loss associated with the reduction is an interaction effect. For a free scalar on $M_4\times S^1$, translation invariance along the circle conserves the KK momentum $n$ exactly. The free action therefore decomposes into uncoupled four-dimensional sectors,
\begin{equation}
S_{\rm free}=\frac12\sum_{n\in\mathbb Z}\int \dd^4x\left[|\partial\phi_n|^2-\frac{n^2}{R^2}|\phi_n|^2\right].
\end{equation}
The Hamiltonian is then a sum of commuting mode Hamiltonians, $H=\sum_n H_n$, and the ground state of a sum of commuting Hamiltonians is the product of their ground states,
\begin{equation}
|\Omega\rangle=\bigotimes_{n\in\mathbb Z}|0_n\rangle .
\end{equation}
Tracing out the tower therefore leaves the zero mode in the pure state $|0_0\rangle$. Its entanglement entropy and its mutual information with the tower both vanish.

A free KK tower carries a nonzero Casimir energy, obtained by summing the zero-point energies of the individual levels \cite{Appelquist:1982zs,Candelas:1983ae}, but its entanglement entropy across levels is zero. Vacuum energy and vacuum entanglement are therefore independent properties of a compactification. Once interactions are switched on, the tower couples to the zero mode. Any nonzero entanglement generated by the reduction therefore first appears at order $\lambda^2$; the remainder of this paper computes that effect at leading order.

\section{Setup and the reduced state at order $\lambda^2$}
\label{sec:setup}

\subsection{The model and the bipartition}
\label{sec:model}

We consider a real scalar field $\Phi(x,y)$ in five dimensions,
\begin{equation}
S=\int \dd^4x\,\dd y\left[\frac12(\partial\Phi)^2-\frac{\lambda_5}{4!}\Phi^4\right],
\qquad y\sim y+2\pi R,\qquad [\lambda_5]=\text{mass}^{-1},
\end{equation}
where the fifth dimension is compactified on a circle of radius $R$. Substituting the mode expansion~\eqref{eq:modeexpansion} and integrating over the compact coordinate yields the four-dimensional interaction Hamiltonian,
\begin{equation}
H_I=\frac{\lambda}{4!}\int \dd^3x\!\!\sum_{n_1+n_2+n_3+n_4=0}\!\!\phi_{n_1}\phi_{n_2}\phi_{n_3}\phi_{n_4},
\qquad \lambda\equiv\lambda_4=\frac{\lambda_5}{2\pi R}.
\label{eq:HI}
\end{equation}
This interaction preserves the KK-number selection rule and couples zero modes to tower states only in pairs or larger groups, as discussed in Section~\ref{sec:kkprimer}. While a single scalar field is a simplification, it serves as a minimal model that captures the correct dimensional scaling and ultraviolet structure of the compactification. Since the entanglement kernel is a sum of non-negative terms, the correlation computed in this sector is, at leading order, a lower bound for any extension that leaves its couplings unchanged, including supersymmetric extensions (Sections~\ref{sec:positivity} and~\ref{sec:evsi}).

To quantify the information lost in dimensional reduction, we partition the Hilbert space into an accessible zero-mode sector and a discarded environment:
\begin{equation}
\mathcal H=\mathcal H_A\otimes\mathcal H_B,\qquad
\mathcal H_A=\{\text{zero-mode quanta with }|\mathbf k|\le\mu\},\qquad
\mathcal H_B=\{\text{everything else}\}.
\end{equation}
The scale $\mu$ represents the energy resolution of the four-dimensional observer, naturally chosen as $\mu\sim1/R$.  $\mathcal H_B$ contains the entire massive tower ($n\neq0$) as well as the zero modes above the resolution scale ($n=0$, $|\mathbf k|>\mu$). These zero modes live in four dimensions, but with $\mu\sim1/R$ their energy exceeds the compactification scale; like the tower, they lie outside the four-dimensional effective description.

A well-defined von Neumann entropy requires a strict tensor factorisation, which in turn necessitates a sharp cutoff in momentum space. Unlike sharp position-space partitions, a sharp momentum cutoff does not introduce entangling-surface divergences.
As shown in Section~\ref{sec:largen}, the choice of $\mu$ and of the shape of the accessible region affects only the overall coefficient and not the dependence on the Kaluza--Klein level. The cut $|\mathbf k|\le\mu$ is imposed in a chosen inertial frame and is not itself boost invariant. The vacuum, however, is boost invariant, since compactification on a circle preserves Lorentz invariance in the four noncompact dimensions, so the same cut imposed in any other inertial frame yields the same entropies for every quantity that does not depend on the ultraviolet cutoff.

\subsection{Reduced density matrix and its spectrum}
\label{sec:bmv}

We follow the perturbative momentum-space entanglement method developed by Balasubramanian, McDermott, and Van Raamsdonk (\text{BMV}) \cite{Balasubramanian:2011wt}. 
In the absence of interactions ($\lambda = 0$), exact Kaluza--Klein momentum conservation keeps the zero mode completely decoupled from the heavy tower, rendering the free vacuum an unentangled product state across all modes. Switching on the 5D interaction $H_I$ creates quantum correlations across sectors. To leading order in perturbation theory, the interacting vacuum $|\Omega\rangle$ receives first-order corrections from intermediate multiparticle excited states $|X\rangle$:
\begin{equation}
|\Omega\rangle = \mathcal N \left[ |0\rangle - \sum_{X \neq 0} |X\rangle \frac{\langle X|H_I|0\rangle}{E_X} \right]
\equiv \mathcal N \left[ |0\rangle + \sum_{X \neq 0} C_X |X\rangle \right],
\qquad
\mathcal N^2 = \left( 1 + \sum_{X \neq 0} |C_X|^2 \right)^{-1},
\end{equation}
where $|0\rangle = |0\rangle_A \otimes |0\rangle_B$ is the free 5D vacuum, $E_X > 0$ is the unperturbed energy of the excited state $|X\rangle$, and $C_X \equiv -\langle X|H_I|0\rangle / E_X = \mathcal{O}(\lambda)$ represents the transition amplitude. 

To compute the entanglement shared between the low-energy subsystem $\mathcal{H}_A$ and the unobserved environment $\mathcal{H}_B$, we categorise the states $|X\rangle$ into pure-$A$ states $|Y\rangle_A|0\rangle_B$, pure-$B$ states $|0\rangle_A|Z\rangle_B$, and mixed states $|Y\rangle_A|Z\rangle_B$ with $Y,Z\neq0$. Here $Y$ and $Z$ label orthonormal multiparticle Fock states of $\mathcal H_A$ and $\mathcal H_B$, specified by the momenta and KK levels. In the finite box of Appendix~\ref{app:norm} these states are discrete, so the sums over them are well defined, and they become phase-space integrals in the continuum limit.

The vacuum state of the interacting theory can then be written as
\begin{equation}
|\Omega\rangle = \mathcal N \left[ |\alpha\rangle_A |0\rangle_B + \sum_{Z \neq 0} |\beta_Z\rangle_A |Z\rangle_B \right],
\end{equation}
where the unnormalised state vectors in $\mathcal{H}_A$ are defined as:
\begin{equation}
|\alpha\rangle = |0\rangle_A + \sum_Y C_{Y0} |Y\rangle_A, \qquad |\beta_Z\rangle = C_{0Z} |0\rangle_A + \sum_Y C_{YZ} |Y\rangle_A.
\end{equation}
Tracing over $B$ and using the orthonormality of $|Z\rangle$,
\begin{equation}
\rho_A = \text{Tr}_B |\Omega\rangle\langle\Omega| 
= \mathcal N^2 \left[ |\alpha\rangle\langle\alpha| + \sum_{Z \neq 0} |\beta_Z\rangle\langle\beta_Z| \right].
\end{equation}

At this order neither pure-$A$ nor pure-$B$ processes generate entanglement entropy (Appendix~\ref{app:eigen}). The entanglement entropy is generated exclusively by genuinely mixed states containing excitations in both sectors ($Y \neq 0$ and $Z \neq 0$).

To order $\lambda^2$, the spectrum of the reduced density matrix $\rho_A$ consists of a single large eigenvalue $p_0\approx1$ associated with the vacuum state, together with small eigenvalues $p_i=\lambda^2\gamma_i$, where $\gamma_i\ge0$ are the eigenvalues of the Gram matrix $\hat G_{YY'}=\lambda^{-2}\sum_{Z\neq0}C_{YZ}C^\ast_{Y'Z}$ (Appendix~\ref{app:eigen}):
\begin{equation}
\begin{gathered}
p_0=1-\lambda^2\Sigma+\mathcal O(\lambda^4),
\qquad
p_i=\lambda^2\gamma_i,
\\
\Sigma\equiv\sum_i\gamma_i=\Tr\hat G=\sum_Y g_Y,
\qquad
g_Y\equiv\hat G_{YY}=\frac{1}{\lambda^2}\sum_{Z\neq0}|C_{YZ}|^2 .
\end{gathered}
\label{eq:spectrum}
\end{equation}
Here, the dimensionless quantity $\Sigma$ represents the trace of a positive-semidefinite Gram matrix $\hat{G}$ formed by the entangled amplitudes (Appendix~\ref{app:eigen}). Since $C_{YZ}$ is the amplitude $C_X$ of the state $|X\rangle=|Y\rangle_A|Z\rangle_B$, the trace is a sum over the states with excitations in both sectors,
\begin{equation}
\Sigma=\frac{1}{\lambda^2}\sum_{\substack{Y\neq0\\Z\neq0}}\frac{|\langle Y,Z|H_I|0\rangle|^2}{E_{Y,Z}^2}.
\end{equation}
Because a trace is basis independent, $\Sigma$ can be evaluated as a phase-space integral without diagonalising $\hat G$. This is what makes the subsequent computation possible.

\subsection{Purity deficit and von Neumann entropy}
\label{sec:twoentropies}

From the spectrum \eqref{eq:spectrum} we can calculate useful diagnostic measures of the reduced state. 

The purity deficit follows directly from the spectrum. Writing the small eigenvalues as $p_i=\lambda^2\gamma_i$, with $\gamma_i\ge0$, one finds (Appendix~\ref{app:eigen})
\begin{equation}
\Tr\rho_A^2=p_0^2+\sum_i p_i^2
=1-2\lambda^2\Sigma+\mathcal O(\lambda^4),
\qquad
1-\Tr\rho_A^2=2\lambda^2\Sigma+\mathcal O(\lambda^4).
\label{eq:purity}
\end{equation}
The second R\'enyi entropy is
\begin{equation}
S_2=-\ln\Tr\rho_A^2=2\lambda^2\Sigma+\mathcal O(\lambda^4).
\end{equation}
This quantity is analytic in $\lambda$ and depends on the eigenvalues of $\hat G$ only through their trace. It is therefore independent of how the continuum of small eigenvalues is binned. Within the present calculation, the purity deficit is the least ambiguous perturbative observable, and it is the quantity used as a perturbative breakdown diagnostic in Ref.~\cite{DuasoPueyo:2024usw}.

The von Neumann entropy is more sensitive to the detailed spectrum. Using $p_0=1-\lambda^2\Sigma+\mathcal O(\lambda^4)$ and $p_i=\lambda^2\gamma_i$, one obtains
\begin{equation}
S_{\rm vN}
=-\Tr(\rho_A\ln\rho_A)
=-p_0\ln p_0-\sum_ip_i\ln p_i
=\lambda^2\Sigma\ln\frac{1}{\lambda^2}
+\lambda^2\left[\Sigma+\sum_i\gamma_i\ln\frac{1}{\gamma_i}\right]
+\mathcal O(\lambda^4).
\label{eq:Sderivation}
\end{equation}
The coefficient of the leading logarithm, $\lambda^2\ln(1/\lambda^2)$, is again the basis-independent quantity $\Sigma=\Tr\hat G$. The accompanying term, however, depends on the individual eigenvalues $\gamma_i$. Eigenvalues do not depend on the choice of basis, but the individual eigenvalues of an operator on a continuum do depend on how the continuum is discretised. As in any perturbative calculation of momentum-space entanglement \cite{Balasubramanian:2011wt}, the purity deficit and the leading logarithm of the von Neumann entropy are therefore independent of the scheme, while the subleading term of the von Neumann entropy is not.

\subsection{Volume law and the intensive kernel}
\label{sec:volumelaw}

Dividing by the spatial volume $V$ gives
\begin{equation}
\frac{S_{\rm vN}}{V}
=\lambda^2\left[\ln\frac{1}{\lambda^2}+\mathcal O(1)\right]\mathcal I+\mathcal O(\lambda^4),
\qquad
\frac{S_2}{V}=2\lambda^2\mathcal I+\mathcal O(\lambda^4),
\label{eq:Sleadinglog}
\end{equation}
with the entanglement kernel
\begin{equation}
\mathcal I\equiv\frac{\Sigma}{V}
=\frac{1}{V}\!\!\sum_{\substack{Y\neq0\\Z\neq0}}\!\!
\frac{|\langle Y,Z|H_I|0\rangle|^2}{\lambda^2E_{Y,Z}^2}.
\label{eq:kerneldef}
\end{equation}
As shown in Appendix~\ref{app:norm}, every three-momentum-conserving vertex in $H_I$ produces exactly one factor of $V$, which cancels the $1/V$ prefactor, so $\mathcal I$ is volume independent and $S\propto V$. Momentum-space entanglement therefore obeys a volume law, in contrast with the area law of position-space entanglement across a spatial boundary \cite{Srednicki:1993im,Eisert:2008ur}. This is expected, since the two subsystems are not spatially localised and there is no entangling surface. The expansion of Section~\ref{sec:bmv} requires $\lambda^2\Sigma=\lambda^2V\mathcal I\ll1$, so the calculation is carried out in a finite box, as in Ref.~\cite{Balasubramanian:2011wt}. In a larger volume $p_0$ is no longer close to one; we assume, as is standard, that the extensive contributions exponentiate, so that the entropy density at order $\lambda^2$ is still controlled by $\mathcal I$, and we quote only densities. The impurity per unit volume of the effective vacuum is therefore an intensive property of the compactification rather than an artefact of the size of the region under consideration.

\subsection{Which vertices can entangle: channel decomposition}
\label{sec:channels}

By Eq.~\eqref{eq:kerneldef}, only states with excitations in both sectors, $Y\neq0$ and $Z\neq0$, contribute to the kernel. We therefore need the vertices of Eq.~\eqref{eq:HI} that place at least one quantum in $\mathcal H_A$ and at least one in $\mathcal H_B$. Recall that $\mathcal H_A$ contains the zero-mode quanta with $|\mathbf k|\le\mu$, which we call soft, while $\mathcal H_B$ contains the entire tower ($n\neq0$, any momentum) together with the zero-mode quanta above the resolution scale ($n=0$, $|\mathbf k|>\mu$). We call the quanta in $\mathcal H_B$ hard.

Kaluza--Klein number conservation, $\sum_in_i=0$, allows zero, two, three or four tower quanta at a vertex, but never exactly one. A vertex with all four quanta soft produces a pure-$A$ state and does not contribute (Appendix~\ref{app:eigen}), and a vertex with all four quanta in $\mathcal H_B$ leaves $\mathcal H_A$ unexcited. The entangling vertices are those with at least one soft quantum and at least one quantum in $\mathcal H_B$. They fall into two groups.

\paragraph{Zero modes only.} These are the $\phi_0^4$ vertices, identical to those of the four-dimensional massless $\lambda\phi^4$ theory studied in Ref.~\cite{Balasubramanian:2011wt}.
We do not recompute them, since they do not contribute to the mutual information with a Kaluza--Klein level, the quantity computed in the rest of the paper (Section~\ref{sec:mutual}).

\paragraph{With tower quanta.} Three configurations involve the tower:
\begin{align}
(2,2):&\quad \phi_0\phi_0\,\phi_n\phi_{-n},\qquad n\ge1,\nonumber\\
(1,3):&\quad \phi_0\,\phi_{n_1}\phi_{n_2}\phi_{n_3},\qquad n_i\neq0,\ \textstyle\sum_in_i=0,\nonumber\\
\text{mixed}:&\quad \phi_0^{\rm soft}\,\phi_0^{\rm hard}\,\phi_n\phi_{-n},\qquad n\ge1.\nonumber
\end{align}
In the $(2,2)$ channel two soft quanta are produced together with a tower pair and in the $(1,3)$ channel one soft quantum is produced with three tower quanta. In the mixed channel one soft quantum is produced with a hard zero mode and a tower pair. The labels count the quanta in $\mathcal H_A$ and $\mathcal H_B$, so by that count the mixed channel is also of type $(1,3)$. Nevertheless we reserve the label $(1,3)$ for three tower quanta. These three channels carry the dependence on the compactification.

The derivation in Appendix~\ref{app:norm} gives the kernels of the three channels,
\begin{align}
\mathcal I_{22}&=\frac12\sum_{n\ge1}\int_{|\mathbf k_{1,2}|\le\mu}
\frac{\dd^3k_1\,\dd^3k_2\,\dd^3q_1}{(2\pi)^9}\,
\frac{1}{16\,\omega_1\omega_2\Omega_1\Omega_2\,E^2},
\label{eq:I22}\\[4pt]
\mathcal I_{13}&=\frac{1}{3!}\sum_{\substack{n_1+n_2+n_3=0\\ n_i\neq0}}\int_{|\mathbf k|\le\mu}
\frac{\dd^3k\,\dd^3q_1\,\dd^3q_2}{(2\pi)^9}\,
\frac{1}{16\,\omega\,\Omega_1\Omega_2\Omega_3\,E^2},
\label{eq:I13}\\[4pt]
\mathcal I_{\rm mix}&=\sum_{n\ge1}\int_{|\mathbf k|\le\mu}\int_{|\mathbf k_0|>\mu}
\frac{\dd^3k\,\dd^3k_0\,\dd^3q_1}{(2\pi)^9}\,
\frac{1}{16\,\omega\,\omega_0\,\Omega_1\Omega_2\,E^2},
\label{eq:Imix}
\end{align}
where $\omega_i=|\mathbf k_i|$ are the soft zero-mode energies, $\omega_0=|\mathbf k_0|$ is the energy of the hard zero mode, and $\Omega_i=\sqrt{\mathbf q_i^2+n_i^2/R^2}$ are the tower energies. The remaining momentum is fixed by three-momentum conservation: $\mathbf q_2=-(\mathbf k_1+\mathbf k_2+\mathbf q_1)$ in the $(2,2)$ channel, $\mathbf q_3=-(\mathbf k+\mathbf q_1+\mathbf q_2)$ in the $(1,3)$ channel, and $\mathbf q_2=-(\mathbf k+\mathbf k_0+\mathbf q_1)$ in the mixed channel. $E$ is the total free energy of the intermediate state. The prefactors $1/2$ and $1/3!$ are identical-particle symmetry factors; the mixed channel has none, since its two zero-mode quanta lie in disjoint momentum regions and each state is counted once. All three kernels are sums of positive terms, a property used in Section~\ref{sec:positivity}.
\section{Channels of the quartic theory}
\label{sec:cha}
\subsection{The $(2,2)$ channel}
\label{sec:22}

We now evaluate Eq.~\eqref{eq:I22}. Decompose the kernel into a sum over KK levels,
\begin{equation}
\mathcal I_{22}=\sum_{n\ge1}J_n,
\label{eq:I22-sum}
\end{equation}
where $J_n$ is the contribution of level $n$. What we want is the large-$n$ behaviour of $J_n$. This fixes the ultraviolet structure of this channel, and the integral that determines it reappears in the cubic theory (Section~\ref{sec:phi3}).

\subsubsection{Large-$n$ behaviour}
\label{sec:largen}
For high KK levels ($n \gg \mu R$),
\[
|\mathbf k_1|,|\mathbf k_2|\le \mu,
\qquad
 \mu R \ll n.
\]
Because $\mathbf{k}_1, \mathbf{k}_2$ are negligible compared to $\frac{n}{R}$, the hard tower momenta that dominate the integral are of the order $|\mathbf q_1|\sim \frac{n}{R}$. Three-momentum conservation then gives
\begin{equation}
    \mathbf{q}_2 = -\mathbf{q}_1 + \mathcal{O}(\mu) \implies \Omega_1 \approx \Omega_2 \approx \Omega(q) \equiv \sqrt{q^2 + n^2/R^2}, \quad \text{where } q \equiv |\mathbf{q}_1|.
\end{equation}
The total energy of the state is
\begin{equation}
    E = \omega_1 + \omega_2 + \Omega_1 + \Omega_2 = 2\Omega(q) \left[ 1 + \frac{\omega_1 + \omega_2}{2\Omega(q)} \right].
\end{equation}
Since $\frac{\omega_1 + \omega_2}{2\Omega(q)} \le \frac{\mu}{n/R} \ll 1$, we can approximate the energy denominator squared as
\begin{equation}
    E^2 \approx 4\Omega(q)^2 = 4 \left( q^2 + \frac{n^2}{R^2} \right).
\end{equation}

With these approximations the contribution $J_n$ to the kernel for large $n$ is given by 
\[
J_n
\simeq
\frac{1}{2}
\frac{1}{16}
\frac{1}{(2\pi)^9}
\left[
\int_{|\mathbf k|\le\mu}
\frac{d^3k}{|\mathbf k|}
\right]^2
\left[
\int
\frac{d^3q}{4( q^2+n^2/R^2)^2}
\right].
\]
The integral of soft modes evaluates to 
\begin{equation}
\int_{|\mathbf k|\le\mu}
\frac{d^3k}{|\mathbf k|}
=
4\pi\int_0^\mu dk\,k
=
2\pi\mu^2.
\label{eq:softint}
\end{equation}
and the integral of hard modes is

\begin{equation}
\int \frac{d^3q_1}{\Omega_1\Omega_2E^2}
\;\longrightarrow\;
\int_{|\mathbf q|\le q_0}\frac{d^3q}{4\,(q^2+n^2/R^2)^{2}}
=\frac{\pi R}{2n}\left[\arctan\frac{q_0R}{n}-\frac{q_0R/n}{1+q_0^2R^2/n^2}\right]
\;\xrightarrow[q_0\gg n/R]{}\;\frac{\pi^2 R}{4n},
\label{eq:qint}
\end{equation}
where $q_0=(\Lambda_{\rm UV}^2-n^2/R^2)^{1/2}$ is the largest three-momentum at level $n$ allowed by the ultraviolet cutoff $\Lambda_{\rm UV}$ of the higher-dimensional theory. For $q_0\gg n/R$ the integral equals $\pi^2R/4n-\pi/q_0$, up to terms of higher order in $n/q_0R$. The correction due to the cutoff is therefore only of relative order $n/\Lambda_{\rm UV}R$.
Assembling the constants of Eq.~\eqref{eq:I22},
\begin{equation}
J_n\;\simeq\;\frac12\cdot\frac{1}{(2\pi)^9}\cdot\frac{1}{16}\cdot\big(2\pi\mu^2\big)^2\cdot\frac{\pi^2R}{4n}
=\frac12\cdot\frac{1}{(2\pi)^9}\cdot\frac{\pi^4\mu^4R}{16\,n},
\end{equation}
that is,
\begin{equation}
\boxed{\;
J_n\;\xrightarrow[n\gg\mu R]{}\;\frac{A}{n R^3}\Big[1+\mathcal O\big(\tfrac{\mu R}{n}\big)\Big],
\qquad
A=\frac{\pi^4(\mu R)^4}{32\,(2\pi)^9}\simeq1.99\times10^{-7}\,(\mu R)^4 .
\;}
\label{eq:Acoeff}
\end{equation}
The resolution scale $\mu$ enters only through the soft integral, Eq.~\eqref{eq:softint}, and hence only through the overall factor $(\mu R)^4$. The hard integral, Eq.~\eqref{eq:qint}, which carries the entire dependence on $n$, does not involve $\mu$. A different choice of $\mu$, or of the shape of the accessible region, therefore changes the coefficient $A$ but leaves the $1/n$ dependence unchanged.

If the higher-dimensional theory is cut off at a physical scale $\Lambda_{\rm UV}$, for example the five-dimensional Planck or string scale, the maximum KK level is $N\simeq\Lambda_{\rm UV}R$. The level sum then behaves as
\[
\sum_{n=1}^{N}J_n\simeq\frac{A}{R^3}\sum_{n=1}^{N}\frac1n
=\frac{A}{R^3}\left[\ln N+\gamma_E+\mathcal O(N^{-1})\right].
\]
Including the finite-$n$ corrections that are important for the first few levels gives
\begin{equation}
\mathcal I_{22}(\Lambda_{\rm UV})
=\frac{A}{R^3}\ln(\Lambda_{\rm UV}R)
+\frac{B(\mu R)}{R^3},
\qquad
\kappa_{22}\equiv R^3\mathcal I_{22}
=A\ln(\Lambda_{\rm UV}R)+B(\mu R),
\label{eq:I22-log}
\end{equation}
where $B(\mu R)$ is a finite, scheme-dependent constant of order $A$, dominated by the lowest levels. The logarithmic ultraviolet sensitivity is mild since $\ln(\Lambda_{\rm UV}R)$ increases only by $\ln10$ for every order of magnitude in $\Lambda_{\rm UV}$. This is the contribution of the $(2,2)$ channel to the kernel of $S_A$.

This marginal behaviour is specific to a single extra dimension. For an internal manifold $K$ of dimension $\dim K$, Weyl's law gives a KK degeneracy $g_n\sim n^{\dim K-1}$, so the corresponding level sum scales as
\[
\sum_n\frac{g_n}{n}\sim\sum_n n^{\dim K-2}.
\]
The sum is therefore logarithmic for $\dim K=1$, linearly divergent for $\dim K=2$, and more strongly divergent for higher codimension. The fact that the coefficient is calculable is thus a special feature of codimension-one compactification.

\subsubsection{Numerical check}
\label{sec:mc}

We checked these results by evaluating Eq.~\eqref{eq:I22} directly by Monte Carlo integration (Appendix~\ref{app:mc}). The computed $J_n$ approach $A/nR^3$ from below, and a power-law fit to the higher levels gives $J_n\propto n^{-0.97\pm0.02}$, consistent with the $1/n$ tail. Runs at different values of $\mu R$ reproduce the $(\mu R)^4$ scaling of the coefficient.

\subsection{The $(1,3)$ and mixed channels: power divergence of $S_A$}
\label{sec:13}

In the $(1,3)$ and mixed channels a single soft quantum is produced together with three hard quanta. Unlike the $(2,2)$ kernel, which is finite level by level (Section~\ref{sec:22}) and whose level sum grows only logarithmically, the kernels of these channels, Eqs.~\eqref{eq:I13} and~\eqref{eq:Imix}, are power divergent already at fixed Kaluza--Klein levels. 

\subsubsection{The $(1,3)$ channel}
\label{sec:13-13}

Let the three hard momenta scale with a common magnitude $q$. Momentum conservation eliminates one of them, leaving the six-dimensional measure $\dd^3q_1\,\dd^3q_2\sim q^5\,\dd q\times(\text{angles})$, while the integrand scales as
\begin{equation}
\frac{1}{\omega\,\Omega_1\Omega_2\Omega_3\,E^2}\;\sim\;\frac{1}{\omega\,q^3\cdot q^2}=\frac{1}{\omega\,q^5}.
\end{equation}
The hard integral therefore behaves as $\int\dd q$ and diverges linearly in the cutoff $\Lambda_{\rm UV}$; the soft integral is finite and contributes a factor of order $\mu^2$.

The sum over the tower strengthens the divergence. Well above the compactification scale the level sums may be replaced by integrals over the internal momentum, $\sum_n\to R\int\dd p_y$, so that each unconstrained hard line in the diagram acquires a phase-space factor of order $qR$. With two independent level sums,
\begin{equation}
\mathcal I_{13}^{\rm 5D}\sim\int_0^{\Lambda_{\rm UV}}q^2R^2\,\dd q\propto\Lambda_{\rm UV}^3R^2,
\end{equation}
a cubic divergence, as expected from five-dimensional power counting with the dimensionful coupling $\lambda_5$. A Monte Carlo evaluation of Eq.~\eqref{eq:I13}
confirms this behaviour.

\subsection{The mixed channel}
\label{sec:mixed}

In the mixed channel the soft quantum is accompanied by a hard zero mode of momentum $\mathbf k_0$ and a tower pair at levels $\pm n$. The ultraviolet behaviour is controlled by the region in which $|\mathbf k_0|$ and $|\mathbf q_1|$ are both large. Taking both of order $q$, momentum conservation gives that $|\mathbf q_2|$ is also of order $q$, hence $\omega_0$, $\Omega_1$ and $\Omega_2$ all scale as $q$ and
\begin{equation}
\frac{1}{\omega\,\omega_0\,\Omega_1\Omega_2\,E^2}\;\sim\;\frac{1}{\omega\,q^5},
\end{equation}
against the measure $\dd^3k_0\,\dd^3q_1\sim q^5\,\dd q$. At fixed $n$ the hard integral is again linearly divergent, with a soft factor of order $\mu^2$.

The two channels differ in their level sums. The mixed channel has one level sum where the $(1,3)$ channel has two. Replacing it by an integral over the internal momentum supplies a single factor $qR$,
\begin{equation}
\mathcal I_{\rm mix}^{\rm 5D}\sim\int_0^{\Lambda_{\rm UV}}qR\,\dd q\propto\Lambda_{\rm UV}^2R,
\end{equation}
one power of $\Lambda_{\rm UV}R$ below the $(1,3)$ channel.

\subsubsection{Origin and consequence}

These divergences are not a consequence of the compactification. A similar divergence exists for the zero modes with one soft and three hard quanta, identified by BMV for four-dimensional $\lambda\phi^4$ \cite{Balasubramanian:2011wt}. In both these cases one mode lies below the resolution scale and three above it. They do not come from a loop and are not removed by renormalising the parameters of the theory: at this order $S_A$ is a sum of squared tree-level amplitudes, and the divergence comes from the phase space of the hard quanta. $S_A$ therefore depends on the ultraviolet cutoff through power-law terms that the low-energy theory does not determine.

\section{Mutual information with a single Kaluza--Klein level}
\label{sec:mutual}

Section~\ref{sec:13} showed that the leading perturbative expression for $S_A$ diverges. In this section we define the quantity computed instead, give its perturbative form, and evaluate it in the quartic theory.

\subsection{Definition}
\label{sec:inclusive}
The entropy $S_A$ measures the correlation of the accessible sector with the entire environment. Following Ref.~\cite{Balasubramanian:2011wt}, we instead compute its correlation with a specified part of the environment, namely the mutual information between the accessible sector and a single Kaluza--Klein level,
\begin{equation}
I_n\;\equiv\;I(A:B_n)=S_A+S_{B_n}-S_{A\cup B_n},
\label{eq:In-def}
\end{equation}
where $B_n$ is the subsystem whose Hilbert space $\mathcal H_{B_n}$ is the Fock space of the modes $\phi_{\pm n}(\mathbf q)$ at Kaluza--Klein level $\pm n$. The ultraviolet completion of the five-dimensional theory is not specified. For the purpose of this calculation, we treat it as a quantum field theory with a cutoff $\Lambda_{\rm UV}$ on the five-dimensional momentum of each quantum, $\mathbf q^2+m_n^2\le\Lambda_{\rm UV}^2$, imposed in the same inertial frame as the resolution cut. This bounds the level number, $|n|\le N\simeq\Lambda_{\rm UV}R$, and the three-momenta in $B_n$, $|\mathbf q|\le(\Lambda_{\rm UV}^2-m_n^2)^{1/2}$. The mutual information compares the reduced density matrix $\rho_{A\cup B_n}$ of the two subsystems with the tensor product $\rho_A\otimes\rho_{B_n}$, which describes the same subsystems with all correlations between them removed. It is the relative entropy
\begin{equation}
I(A:B)=S\big(\rho_{AB}\,\big\|\,\rho_A\otimes\rho_B\big),
\label{eq:mi-is-rel}
\end{equation}
and vanishes exactly when $\rho_{A\cup B_n}=\rho_A\otimes\rho_{B_n}$. Mutual information measures the total correlation, classical and quantum, between the two subsystems; for free fields it is zero (Section~\ref{sec:free}). Since $A\cup B_n$ is not the whole system, $\rho_{A\cup B_n}$ is mixed.

At leading order the entropy $S_A$ of any subsystem $A$ is a sum over the excited states $|X\rangle$ of Section~\ref{sec:bmv} that contain at least one quantum in $A$ and at least one in its complement; states with all their quanta on one side do not contribute (Appendix~\ref{app:eigen}). Consider a state that excites $A$ and the remainder of the environment but not $B_n$. It contributes to $S_A$ and to $S_{A\cup B_n}$ but not to $S_{B_n}$, so its contributions to $I_n$ cancel; the same holds for a state that excites $B_n$ and the remainder but not $A$. Only states that excite both $A$ and $B_n$ contribute to $I_n$. The zero-mode-only configurations of Section~\ref{sec:channels}, including the one whose divergence was found in Ref.~\cite{Balasubramanian:2011wt}, contain no Kaluza--Klein quantum and therefore drop out. The three channels involving the tower contribute provided one of their tower quanta lies at level $\pm n$.

\subsection{Perturbative form and additivity}
\label{sec:In-form}
\label{sec:convadd}

Write the Hilbert space as $\mathcal H_A\otimes\mathcal H_{B_n}\otimes\mathcal H_C$, where $\mathcal H_C$ contains all modes outside $A$ and $B_n$: the zero modes above the resolution scale and all levels other than $\pm n$. For a pure state, the leading-order mutual information between $A$ and $B_n$ is a sum over the squared amplitudes entering Eq.~\eqref{eq:Sleadinglog}, each weighted by zero unless it excites both $A$ and $B_n$, and otherwise by two if $C$ is unexcited and by one if it is excited \cite{Balasubramanian:2011wt}. Writing $J_n^{(c)}$ for the contribution of level $n$ to the kernel of channel $c$ and $w_c$ for its weight,
\begin{equation}
I_n=\lambda^2\Big[\ln\tfrac{1}{\lambda^2}+\mathcal O(1)\Big]\,V\mathcal J_n,
\qquad
I_{2,n}=2\lambda^2V\mathcal J_n,
\qquad
\mathcal J_n=\sum_c w_c\,J_n^{(c)} .
\label{eq:In-J}
\end{equation}
The first expression is the von Neumann mutual information. Its leading logarithm is independent of the scheme, while the $\mathcal O(\lambda^2)$ remainder depends on the individual small eigenvalues, and hence on how the continuum is discretised (Section~\ref{sec:twoentropies}). The second is the R\'enyi-2 mutual information, which is complete at order $\lambda^2$ and has no such remainder. Unlike the von Neumann mutual information, the R\'enyi-2 combination is not a relative entropy and need not be positive in general; at this order it is positive because $\mathcal J_n$ is a sum of non-negative terms. The R\'enyi-2 mutual information and the leading logarithm of the von Neumann one are proportional to the weighted kernel $\mathcal J_n$ and share its dependence on $n$.

At leading order, mutual informations with different levels are exactly additive provided no state excites all four of $A$, $B_n$, $B_m$ ($m\neq n$) and the complement simultaneously \cite{Balasubramanian:2011wt}. For the contribution of any channel whose states excite at most one Kaluza--Klein level, therefore,
\begin{equation}
I\Big(A:\bigcup_{1\le n\le N}B_n\Big)=\sum_{n=1}^{N}I_n
\label{eq:additivity}
\end{equation}
holds exactly at order $\lambda^2$ for the R\'enyi-2 mutual information and for the leading logarithm of the von Neumann one, and the level sum is itself a mutual information.

\subsection{The quartic theory}
\label{sec:whichchannel}

In the quartic theory, three channels contribute to $I_n$ (Section~\ref{sec:channels}). A $(2,2)$ state excites only $A$ and $B_n$ and carries a weight of two. The mixed channel excites the hard zero mode, and the $(1,3)$ channel excites spectator levels, both of which lie in the complement. These channels thus carry a weight of one. The weighted kernel is therefore
\begin{equation}
\mathcal J_n=2J_n^{(2,2)}+J_n^{\rm mix}+J_n^{(1,3)},
\label{eq:Jquartic}
\end{equation}
where $J_n^{(2,2)}$ is the per-level kernel of Section~\ref{sec:22}, and $J_n^{\rm mix}$ and $J_n^{(1,3)}$ are the contributions of level $n$ to Eqs.~\eqref{eq:Imix} and~\eqref{eq:I13}. Numerical values below are taken from the Monte Carlo evaluation of Appendix~\ref{app:mc}.

\paragraph{The $(2,2)$ and mixed channels.} Both channels arise from the vertex $\phi_0\phi_0\phi_n\phi_{-n}$, and Eqs.~\eqref{eq:I22} and~\eqref{eq:Imix} have the same integrand. They differ only in the momentum $\mathbf k_0$ of the second zero-mode quantum, which is soft in the $(2,2)$ channel and hard in the mixed channel. The factor $1/2$ of the $(2,2)$ kernel counts each pair of soft quanta once, and the weight two restores both orderings. The weighted sum is therefore a single integral in which $\mathbf k_0$ ranges over its entire domain,
\begin{equation}
2J_n^{(2,2)}+J_n^{\rm mix}\;=\;K_n\;\equiv\;\int_{|\mathbf k|\le\mu}\int_{|\mathbf k_0|\le\Lambda_{\rm UV}}
\frac{\dd^3k\,\dd^3k_0\,\dd^3q_1}{(2\pi)^9}\,
\frac{1}{16\,\omega\,\omega_0\,\Omega_1\Omega_2\,E^2},
\label{eq:Kn}
\end{equation}
with $\mathbf q_2=-(\mathbf k+\mathbf k_0+\mathbf q_1)$. The resolution scale no longer bounds $\mathbf k_0$. In the $(2,2)$ channel the restriction $|\mathbf k_0|\le\mu$ confines the second zero-mode quantum to a region small compared to $m_n$, so that the tower pair carries momenta of the order of $m_n$. This is why we obtain the term $A/n$ of Section~\ref{sec:largen}. In $K_n$ the second mode has momenta all the way up to the cutoff, and the integral is dominated by $\mathbf k_0$ and $\mathbf q_1$ of comparable size, far above $\mu$.

For $m_n\gg\mu$ the soft momentum $\mathbf k$ can be neglected in $\Omega_2$ and $E$, and $K_n$ becomes $[2\pi\mu^2/16(2\pi)^9]\,F(m_n)$, where
\begin{equation}
F(m)=\int_{|\mathbf k_0|\le\Lambda_{\rm UV}}\dd^3k_0\int\dd^3q_1\,
\frac{1}{\omega_0\,\Omega_1\Omega_2\,(\omega_0+\Omega_1+\Omega_2)^2},
\label{eq:Fm}
\end{equation}
with $\Omega_1=(\mathbf q_1^2+m^2)^{1/2}$ and $\Omega_2=[(\mathbf k_0+\mathbf q_1)^2+m^2]^{1/2}$. Measuring all momenta in units of $m$ separates the dimensionful factors,
\begin{equation}
F(m)=m\,\Phi\!\left(\frac{\Lambda_{\rm UV}}{m}\right),
\label{eq:Phi}
\end{equation}
where $\Phi(x)$ is the same integral evaluated at unit mass with the cutoff at $x$. Evaluated numerically, $\Phi$ grows linearly at large $x$,
\begin{equation}
\Phi(x)=a'x-\beta+\mathcal O(1/x),\qquad \beta\simeq82.5 .
\label{eq:beta}
\end{equation}
where the slope $a'$ depends on the shape of the cutoff and the intercept $\beta$ does not. The linear term is the growth at fixed level found for the mixed channel in Section~\ref{sec:mixed}. Summing over the $N\simeq\Lambda_{\rm UV}R$ levels we get back the $\Lambda_{\rm UV}^2$ growth found there. For the cutoff of Section~\ref{sec:inclusive}, $a'\simeq22.4$. Inserting Eq.~\eqref{eq:beta} into Eq.~\eqref{eq:Phi} with $m=m_n=n/R$ gives
\begin{equation}
K_n\;\simeq\;\frac{(\mu R)^2}{R^3}\Big[a\,\Lambda_{\rm UV}R-b\,n\Big],
\qquad \mu R\ll n\ll\Lambda_{\rm UV}R,
\label{eq:Knlin}
\end{equation}
with $a=2\pi a'/16(2\pi)^9\simeq5.76\times10^{-7}$ and $b=2\pi\beta/16(2\pi)^9\simeq2.1\times10^{-6}$. 
The Monte Carlo evaluation of Eq.~\eqref{eq:Kn} (Appendix~\ref{app:mc}) agrees with Eq.~\eqref{eq:Knlin} to about $1\%$ for $n\lesssim0.05\,\Lambda_{\rm UV}R$.

\paragraph{The $(1,3)$ channel.} For fixed levels the $(1,3)$ integral has the same structure as $F$.
At fixed external level $n$ one spectator level is summed over, the other being fixed by $n_3=-n-n_2$, and the spectator levels run up to $\Lambda_{\rm UV}R$. The sum over spectator levels multiplies both terms of $F$ by $\Lambda_{\rm UV}R$. The linear growth becomes quadratic, and the mass of the quantum at level $n$ gives a term of order $\Lambda_{\rm UV}m_n$. The Monte Carlo evaluation (Appendix~\ref{app:mc}) confirms this form,
\begin{equation}
J_n^{(1,3)}\;\simeq\;\frac{(\mu R)^2}{R^3}\Big[c\,(\Lambda_{\rm UV}R)^2-c'\,\Lambda_{\rm UV}R\,n\Big],
\qquad n\ll\Lambda_{\rm UV}R,
\label{eq:J13lin}
\end{equation}
with $c\approx3\times10^{-7}$ and $c'\approx4\times10^{-7}$, both of which depend on the shape of the cutoff. Summing the first term over the $N\simeq\Lambda_{\rm UV}R$ levels gives the growth of $\mu^2R^2\Lambda_{\rm UV}^3$ derived in Section~\ref{sec:13-13}.

\paragraph{Consequence.} Combining Eqs.~\eqref{eq:Jquartic}--\eqref{eq:J13lin},
\begin{equation}
R^3\mathcal J_n\;\simeq\;(\mu R)^2\Big[c\,(\Lambda_{\rm UV}R)^2+a\,\Lambda_{\rm UV}R-\big(c'\,\Lambda_{\rm UV}R+b\big)\,n\Big].
\label{eq:wholetotal}
\end{equation}
The mutual information of the quartic theory depends on the cutoff, and it decreases linearly with $m_n$, with a slope that grows with $\Lambda_{\rm UV}$, instead of falling as a power of the level mass.
 Since the five-dimensional quartic coupling has $[\lambda_5]=-1$, the theory is valid only up to a finite scale, and $\Lambda_{\rm UV}$ corresponds to the scale at which the unknown ultraviolet completion takes over (Section~\ref{sec:running}). The cutoff dependence of Eq.~\eqref{eq:wholetotal} is therefore not an artefact of the regularisation. It shows that in the quartic theory the correlation of the accessible sector with a Kaluza--Klein level is set by the physics at the cutoff.

In the quartic theory the $(2,2)$ channel contributes $A\ln N$ to the kernel of $S_A$, Eq.~\eqref{eq:I22-log}. In the level-summed mutual information, the $(2,2)$ and mixed channels instead combine into $\sum_nK_n$, which grows as $\mu^2R\Lambda_{\rm UV}^2$ (Section~\ref{sec:mixed}), while the $(1,3)$ channel grows as $\mu^2R^2\Lambda_{\rm UV}^3$. The level-summed information of the quartic theory is therefore controlled by the cutoff.

\section{The cubic theory}
\label{sec:phi3}

\subsection{Channels}
\label{sec:phi3-channels}

Consider the same compactification with a cubic self-interaction,
\begin{equation}
S=\int \dd^4x\,\dd y\left[\frac12(\partial\Phi)^2-\frac{\lambda_5}{3!}\Phi^3\right],
\qquad [\lambda_5]=+\tfrac12 ,
\end{equation}
whose four-dimensional coupling, $\lambda_3\equiv\lambda_5/(2\pi R)^{1/2}$, has mass dimension one. The five-dimensional coupling has positive mass dimension, so the theory is super-renormalisable. Each vertex has three legs, and Kaluza--Klein number conservation admits three classes of vertices: $(0,0,0)$, $(0,n,-n)$, and $(n,m,-n-m)$ with all legs in the tower.

Only the second class contributes to $I_n$. The first contains no Kaluza--Klein quantum and has no excitations in $B_n$. The third contains no zero-mode quantum and leaves the accessible sector unexcited. Thus, the channel of interest for our purpose is one in which a soft quantum of the accessible sector is produced together with a tower pair at levels $\pm n$.
When the zero-mode quantum of the second class is hard, the state does not excite $A$ and does not contribute either. There is no analogue of the mixed or $(1,3)$ channels in the cubic theory. Hence the weighted kernel of Eq.~\eqref{eq:In-J} is $\mathcal J_n=2J_n^{(3)}$, where $J_n^{(3)}$ denotes the contribution of level $n$ to the kernel of this channel.

\subsection{Ultraviolet finiteness}
\label{sec:phi3-finite}
For the $(0,n,-n)$ channel, the momentum constraint leaves two free momenta, one of which is the soft momentum bounded by $\mu$. Thus, only one hard momentum is free. At large $q$ one has $\Omega_1\simeq\Omega_2\simeq q$ and $E\simeq2q$, and
\begin{equation}
\int\frac{\dd^3q}{\Omega_1\Omega_2E^2}\;\longrightarrow\;\int_{|\mathbf q|\le q_0}\frac{\dd^3q}{4(q^2+m_n^2)^2}\;\xrightarrow[q_0\gg m_n]{}\;\frac{\pi^2R}{4n},
\end{equation}
the same integral as in Eq.~\eqref{eq:qint}. In the quartic theory, the fourth leg supplies a second free hard momentum, which is the origin of the power divergences; with three legs there is no such cutoff dependence. In the cubic theory $I_n$ is therefore finite, and the cutoff affects it only through corrections suppressed by $m_n/\Lambda_{\rm UV}$ (Eq.~\eqref{eq:qint}). The zero-mode configurations are those of the four-dimensional cubic theory studied by BMV \cite{Balasubramanian:2011wt}, and they do not contribute to $I_n$.

\subsection{Result}
\label{sec:phi3-result}

Assembling the constants as in Section~\ref{sec:largen}, with a single soft integral $\int_{|\mathbf k|\le\mu}\dd^3k/|\mathbf k|=2\pi\mu^2$ instead of its square,
\begin{equation}
\boxed{\;
J_n^{(3)}\;\xrightarrow[n\gg\mu R]{}\;\frac{A_3}{n\,R},
\qquad
A_3=\frac{\pi^3(\mu R)^2}{16\,(2\pi)^6}=\frac{(\mu R)^2}{1024\,\pi^3}\simeq3.15\times10^{-5}\,(\mu R)^2 .
\;}
\label{eq:A3}
\end{equation}
The Monte Carlo evaluation (Appendix~\ref{app:mc}) confirms this result. Our numerical computations show that for $\mu R=1$, $nRJ_n^{(3)}$ approaches $A_3$ from below, lies within $1\%$ of it for $n\gtrsim30$, and is about $25\%$ smaller at $n=1$, where $m_n$ is comparable to $\mu$. From Eq.~\eqref{eq:In-J} with $\mathcal J_n=2J_n^{(3)}$, the mutual information with level $n$ is 
\begin{equation}
I_n=2\lambda_3^2\Big[\ln\tfrac{1}{\lambda_3^2R^2}+\mathcal O(1)\Big]\,V\,\frac{A_3}{nR},
\qquad
I_{2,n}=4\lambda_3^2\,V\,\frac{A_3}{nR},
\label{eq:phi3In}
\end{equation}
up to corrections of relative order $\mu R/n$. Since $\lambda_3$ has mass dimension one, the dimensionless coupling is $\lambda_3R$, the ratio of $\lambda_3$ to the compactification scale $1/R$. Every state of the channel excites $A$ and $B_n$, so Eq.~\eqref{eq:additivity} holds for the cubic theory. The sum of Eq.~\eqref{eq:phi3In} over levels is the complete mutual information between the accessible sector and the tower (Section~\ref{sec:renorm}).

 The dependence on the level $n$ is due to the hard integral, Eq.~\eqref{eq:qint}, which appears in the cubic channel and the $(2,2)$ channel. Its integrand $q^2/(q^2+m_n^2)^2$ grows as $q^2/m_n^4$ for $q\ll m_n$ and falls as $q^{-2}$ for $q\gg m_n$, so the integral is dominated by momenta $|\mathbf q|\sim m_n$.
 A region of fixed size in momentum space behaves differently. With an upper limit $Q\ll m_n$ for the momenta the integral becomes $\pi Q^3/3m_n^4$,
 which falls as $m_n^{-4}$, like the correlation between modes at separated scales found by BMV \cite{Balasubramanian:2011wt}. A KK level, by contrast, extends over all three-momenta and contains modes with momenta $|\mathbf q|\sim m_n$ that dominate the integral. Its correlation with the accessible sector therefore falls as $m_n^{-4}\times m_n^3=m_n^{-1}$, which is exactly the power at which the sum over levels stops converging and grows logarithmically (Section~\ref{sec:renorm}). The cubic potential is unbounded below, but at this order $I_n$ depends only on the vertex $\phi_0\phi_n\phi_{-n}$, so the result holds in any theory with such a vertex, including theories whose potential is bounded below.

\subsection{Sum over levels and the entropy of the accessible sector}
\label{sec:renorm}

Summing Eq.~\eqref{eq:phi3In} over levels with Eq.~\eqref{eq:additivity}, the mutual information between the accessible sector and the first $N$ levels of the tower in the cubic theory is
\begin{equation}
\boxed{\;
I\Big(A:\bigcup_{n\le N}B_n\Big)
=2\lambda_3^2\Big[\ln\tfrac1{\lambda_3^2R^2}+\mathcal O(1)\Big]\,V\,\frac{A_3}{R}\Big[\ln N+B_3+\mathcal O(N^{-1})\Big],
\;}
\label{eq:sKKfinal}
\end{equation}
with R\'enyi-2 form
\begin{equation}
\boxed{\;
I_2\Big(A:\bigcup_{n\le N}B_n\Big)=4\lambda_3^2\,V\,\frac{A_3}{R}\Big[\ln N+B_3+\mathcal O(N^{-1})\Big].
\;}
\label{eq:s2final}
\end{equation}
Here $B_3$ is a constant of order one, determined by the lowest levels. Since only a single channel contributes to the summation, Eq.~\eqref{eq:s2final} is the complete R\'enyi-2 mutual information at order $\lambda_3^2$, while Eq.~\eqref{eq:sKKfinal} gives the von Neumann mutual information to leading logarithmic accuracy. Both are positive, proportional to the spatial volume, and grow as $\ln N$ as long as the perturbative expansion remains valid (Section~\ref{sec:running}). Since additional fields add only non-negative terms to the kernel at this order (Section~\ref{sec:positivity}), both are lower bounds for any extension, supersymmetric or not, that contains this sector with its couplings unchanged.

 The coefficient $A_3=(\mu R)^2/1024\pi^3$ comes from the phase space integral and depends on the resolution cut $\mu R$ (Section~\ref{sec:largen}). The overall normalisation also involves the $\mathcal O(1)$ term accompanying the perturbative logarithm, which is absent from Eq.~\eqref{eq:s2final}, and the constant $B_3$. The factor $\ln N$ depends only on the extent of the tower and grows without bound as more levels are included. See Section~\ref{sec:running} for physical estimates on the values of $N$.

\paragraph{Entropy of the accessible sector.} In the cubic theory the states that contribute to $S_A$ are those of Eq.~\eqref{eq:sKKfinal}, together with states containing only zero modes. Each of these states has at most one free hard momentum, so the contribution of every level is finite, and
\begin{equation}
S_A=\lambda_3^2\Big[\ln\tfrac{1}{\lambda_3^2R^2}+\mathcal O(1)\Big]\,V\Big[\frac{A_3}{R}\big(\ln N+B_3\big)+D_3\,\mu\Big],
\qquad D_3=\frac{\ln2-\tfrac12}{128\pi^4}.
\label{eq:SAcubic}
\end{equation}
The term $D_3\mu$ is the contribution of the states containing only zero modes. Their integral converges and involves no scale other than $\mu$.\footnote{With $\mathbf k_3=-\mathbf k_1-\mathbf k_2$ and $\omega_i=|\mathbf k_i|$, $D_3\mu=\frac{1}{3!}\frac{1}{8(2\pi)^6}\int\dd^3k_1\,\dd^3k_2\,[\omega_1\omega_2\omega_3(\omega_1+\omega_2+\omega_3)^2]^{-1}$, over configurations with at least one soft and one hard quantum, where $1/3!$ is the symmetry factor of three identical quanta. This integral is the massless limit of $I_3(\mu)$ in Ref.~\cite{Balasubramanian:2011wt}. 
The integral evaluates to $24\pi^2(\ln2-\tfrac12)\mu$, which we have confirmed numerically. This differs slightly from the asymptotic value quoted in Ref.~\cite{Balasubramanian:2011wt}.} This term is the soft--hard entanglement of the four-dimensional cubic theory studied by BMV \cite{Balasubramanian:2011wt}, and it is independent of $R$. The tower term grows as $\ln N$ while this term is fixed, so for a long tower the entropy of the accessible sector is dominated by its correlation with the tower. Through $\ln N$, $S_A$ grows logarithmically with the cutoff, as power counting predicts for the cubic theory in five dimensions.

\section{Energy versus information under supersymmetry}
\label{sec:evsi}

A compactification is usually characterised by its vacuum energy, which supersymmetry cancels. The correlation computed in the preceding sections is a different property of the same vacuum. We show that, at leading order, supersymmetry cannot cancel it, which is the basis of the lower bound stated in Section~\ref{sec:renorm}.

\subsection{Vacuum energy}
\label{sec:vac}

For a single real bosonic degree of freedom on $M_4\times S^1$, the free tower carries the Casimir energy density
\begin{equation}
\rho_{\rm Cas}=-\frac{3\,\zeta(5)}{64\pi^6R^4}
\end{equation}
\cite{Appelquist:1982zs,Candelas:1983ae,Ponton:2001hq}. With unbroken supersymmetry both contributions cancel. The Hamiltonian is a sum of squares of the supercharges,
\begin{equation}
H=\frac{1}{4}\sum_\alpha\{Q_\alpha,Q_\alpha^\dagger\}\ge0,\qquad Q_\alpha|0\rangle=0,
\end{equation}
so the vacuum energy vanishes to all orders in perturbation theory \cite{Zumino:1974bg,Witten:1981nf}.

\subsection{Positivity of the correlation}
\label{sec:positivity}

The correlation behaves differently. At order $\lambda^2$ the entanglement kernel, Eq.~\eqref{eq:kerneldef}, is a sum of non-negative terms, one for each state with excitations in both sectors. The superpartners open new channels, whose states contain fermions and are orthogonal to the bosonic ones. Each channel adds its own non-negative contribution, and no cancellation between channels is possible. Formally, the Gram matrix that controls the purity, Eq.~\eqref{eq:purity}, and the leading logarithm of the entropy, Eq.~\eqref{eq:Sderivation}, is positive semidefinite (Appendix~\ref{app:eigen}), and additional channels can only increase its trace. The same holds for the weighted kernel $\mathcal J_n$ of the mutual information, whose weights are non-negative.

Supersymmetry therefore removes the vacuum energy but leaves the correlation, including its dependence on the cutoff in the quartic theory. At order $\lambda^2$ the correlation computed in the scalar sector is a lower bound for any supersymmetric completion that contains it with its couplings unchanged,
\begin{equation}
\mathcal I^{\rm SUSY}\ \ge\ \mathcal I^{\rm bosonic}.
\end{equation}
The size of the enhancement is model dependent and requires explicit computation. The argument assumes that the completion leaves the amplitudes of the bosonic states unchanged. Additional interactions producing the same bosonic states would interfere with them and are not covered by it. The bound is established at leading order, where the spectrum of $\rho_A$ is controlled by $\Tr\hat G$; whether it survives at higher orders is not examined here.

A supersymmetric Minkowski compactification has exactly vanishing vacuum energy, yet its accessible sector is in a mixed state. A vanishing vacuum energy therefore does not indicate that the four-dimensional description captures the full state.

\section{Implications for the four-dimensional description}
\label{sec:eft}

This section considers what the results imply for the four-dimensional description.

\subsection{Decoupling of amplitudes versus information}
\label{sec:slow}

The low-energy observables of the four-dimensional theory do not depend on how far the tower extends, but the reduced state does. In the cubic theory the information shared with level $n$ falls as $1/m_n$, Eq.~\eqref{eq:phi3In}, whereas, once the mass and couplings of the zero mode are renormalised, the contribution of the same level to a low-energy amplitude is suppressed by $1/m_n^2$. This difference in power separates a convergent sum from a divergent one. The corrections to amplitudes sum to $\sum_n m_n^{-2}=\pi^2R^2/6$, are dominated by the lightest levels, and are reproduced in the effective theory by local operators suppressed by $R^2$. This is the content of the Appelquist--Carazzone theorem \cite{Appelquist:1974tg}, which shows that the deep ultraviolet sector of the tower does not affect low-energy amplitudes. The information sum is instead not dominated by any finite set of levels. Every octave of the tower with $\mu R\ll n\ll\Lambda_{\rm UV}R$ contributes the same amount, proportional to $A_3\ln2$. Doubling the number of levels changes the amplitude corrections by $\mathcal O(1/N)$ but adds a fixed amount to the information. Unlike the corrections to amplitudes, this growth cannot be reproduced by local terms in the action of the zero mode, since such an action gives the zero modes a pure state and contains no correlation with the tower (Section~\ref{sec:local}).

\subsection{The information is not in the local effective action}
\label{sec:local}

Integrating out the tower gives a four-dimensional effective Lagrangian consisting of the zero-mode kinetic and interaction terms together with local higher-dimension operators suppressed by powers of $R$. The coefficients of these operators form the saturated sums of Section~\ref{sec:slow} and are dominated by the first few levels. The mutual information computed above is not contained in this local derivative expansion. The full effective action obtained by integrating out the tower also contains terms nonlocal in time on the scale $\Delta t\sim1/m_n$ \cite{Costa:2022bvs,MartinsCosta:2022bjr}. Terms local in time generate unitary evolution and leave the state of the zero modes pure, so their entanglement with the tower arises entirely from the nonlocal terms. A derivative expansion discards them, although they carry the unsaturated information to which every octave of the tower contributes.

The same conclusion follows from the dependence on the level. A derivative expansion of the contributions of the tower produces only even powers of $1/m_n$, whereas the correlation with a level falls as $1/m_n$. Donath and Pajer found that integrating out heavy fields in the in-in formalism generates contributions odd in $1/M$ that a derivative expansion cannot produce, and associated them with dissipation \cite{Donath:2024utn}. An analogous limitation was found by Green and Sun for equal-time correlators. Matching them between short and long distances requires additional terms, localised at the time of measurement, that the effective action does not contain \cite{Green:2024cmx}. Our result points in a similar direction for the state itself, with one difference: their additional terms correct only local and semi-local contributions, whereas the contribution missing here is spread over the whole tower and grows with the number of levels.

\paragraph{Observables and the state.} The result concerns the state rather than the observables. Every measurement that a low-energy observer performs on a single copy of the system is the expectation value of an operator, $\langle\mathcal O\rangle=\Tr_A(\rho_A\mathcal O)$, which is linear in $\rho_A$, and the local effective theory reproduces such quantities at energies $E\ll1/R$ up to corrections of order $E^2R^2$. 
Entropies, purities and mutual information are instead nonlinear functionals of $\rho_A$, and cannot be written as the expectation value of any operator acting on a single copy of the state. The purity, for example, is the expectation value of the swap operator on two identically prepared copies of the accessible sector, $\Tr\rho_A^2=\Tr[(\rho_A\otimes\rho_A)P_{\rm swap}]$ \cite{Ekert:2002qtj,Daley:2012xhf}. The local effective theory therefore describes low-energy correlation functions accurately, but the low-energy state only incompletely, and the missing information has the tower-enhanced size computed above. The distinction becomes physically consequential when a nonlinear functional of the state is itself the prediction, such as for cosmological decoherence \cite{Brahma:2020zpk,Brahma:2023hki,Burgess:2022nwu,Colas:2024xjy,Salcedo:2024smn}.

\subsection{Operational meaning of the mutual information}
\label{sec:operational}

The mutual information $I_n\equiv I(A:B_n)$ has a direct operational meaning for any observer, or environment, with access to level $n$. It is the relative entropy between the joint state $\rho_{AB_n}$ and the product $\rho_A\otimes\rho_{B_n}$ in which level $n$ is decoupled from the accessible sector. By the quantum Stein lemma \cite{Hiai:1991mxv,Ogawa:1999gv}, it is the optimal rate of asymmetric discrimination between the two states.  Given $k$ copies of the joint state of $A$ and $B_n$, the probability of mistaking the decoupled state for the correlated one falls as $e^{-kI_n}$ while the probability of the opposite error is held fixed.

The mutual information also bounds every connected correlation between an operator on $A$ and one on $B_n$ \cite{Wolf:2007tdq},
\begin{equation}
\frac{\langle\mathcal O_A\mathcal O_{B_n}\rangle_c^2}{2\,\|\mathcal O_A\|^2\,\|\mathcal O_{B_n}\|^2}\;\le\;I_n ,
\end{equation}
for bounded operators $\mathcal O_A$ and $\mathcal O_{B_n}$, and it equals the minimal rate of local randomness needed to erase the correlation between the two sectors \cite{Groisman:2005dbo}.

In the cubic theory these quantities inherit the dependence of $I_n$ on the level. The contribution of a level to low-energy amplitudes falls as $1/m_n^2$, but the rate at which the correlated state can be distinguished from the decoupled one, and the cost of erasing the correlation, fall only as $1/m_n$, and the bound on the connected correlations falls as $m_n^{-1/2}$. 

These statements quantify how weakly, and how slowly, a heavy level decouples from the state. Since $I_n\propto1/m_n$, the number of copies that an observer with access to level $n$ needs to distinguish the correlated state from the decoupled one grows in proportion to $m_n$.
More importantly, dimensional reduction discards the tower collectively. Because the mutual information is additive over levels, the correlation with the first $N$ levels is $I(A:\bigcup_{n\le N}B_n)=C(\ln N+B_3)$ with $C=2\lambda_3^2[\ln(1/\lambda_3^2R^2)+\mathcal O(1)]VA_3/R$, Eq.~\eqref{eq:sKKfinal}, and the probability of mistaking the decoupled state for the correlated one falls as $N^{-kC}$. Whereas the corrections to amplitudes saturate at the lowest levels, the correlated state becomes steadily easier to distinguish from the decoupled one as more levels are included.

These quantities become relevant whenever the tower acts as an environment for the accessible sector, for example, in the laboratory systems discussed in Section~\ref{sec:lab}.

\subsection{Magnitude of the tower contribution and validity of the expansion}
\label{sec:running}

Since the logarithm in Eq.~\eqref{eq:sKKfinal} grows without bound, its size is set by where the tower ends. The gravitational bound on the number of species, $\Lambda_{\rm UV}\le\MPl/\sqrt N$, gives $N\simeq(\MPl/M_{\KK})^{2/3}$ \cite{Dvali:2007hz,Dvali:2007wp}, so that $\ln N$ ranges from about $25$ to about $50$ for compactification scales between a TeV and $10^{-3}$~eV. A string completion supersedes the Kaluza--Klein description at the string scale, which can only lower $N$. In the quartic theory the five-dimensional coupling becomes strong near the KK level $12\pi^2/\lambda$, so that $\ln N\lesssim\ln(12\pi^2/\lambda)$.

The second R\'enyi entropy per accessible mode provides an intensive measure of the mixedness of the reduced state. Normalising by the number of accessible zero modes, $N_A=V\mu^3/(6\pi^2)$, the contribution of the Kaluza--Klein tower in the cubic theory to the second R\'enyi entropy of the accessible sector, which is half the corresponding R\'enyi-2 mutual information, gives
\begin{equation}
\varepsilon\;\equiv\;\frac{S_2}{N_A}\bigg|_{\rm tower}
=\frac{3}{256\pi}\,\frac{(\lambda_3R)^2}{\mu R}\,\ln N .
\label{eq:epsilon}
\end{equation}
For perturbative couplings, $\lambda_3R\lesssim1$, and $\mu R$ of order one, this quantity remains small, and it reaches order unity only when $(\lambda_3R)^2\ln N\simeq270\,\mu R$. By the entanglement-based unitarity criterion of Ref.~\cite{DuasoPueyo:2024usw}, this marks the point at which the leading-order description of the reduced state breaks down. The breakdown is collective: each level remains weakly correlated with the accessible sector, but the accumulation over the tower makes $\lambda_3^2\ln N$ the effective expansion parameter. On an internal manifold with $\dim K\ge2$, Weyl's law turns the logarithmic level sum into a power of $N$, so $\varepsilon$ would reach order unity at parametrically weaker coupling; establishing this conclusively requires the corresponding higher-dimensional kernel, which is not computed here. In the quartic theory the tower contribution is instead dominated by the $(1,3)$ channel (Section~\ref{sec:whichchannel}), and $\varepsilon\propto\lambda^2(\Lambda_{\rm UV}R)^3/\mu R$ with a coefficient that depends on the shape of the cutoff. The leading-order description breaks down at $\Lambda_{\rm UV}R\propto\lambda^{-2/3}$, which at weak coupling lies below the level $\propto\lambda^{-1}$ at which the five-dimensional coupling becomes strong.

\section{Discussion}
\label{sec:discussion}

\paragraph{Derivative couplings.} The $1/m_n$ tail requires the accessible sector to couple to pairs of tower modes without derivatives. The graviton zero mode and gauge zero modes also couple through vertices containing one zero-mode field and two tower fields, with the same field content as the cubic vertex. These vertices come from the kinetic terms rather than from a potential, so they are present for any tower, independent of its self-interactions, but they carry derivatives of the tower fields. Specifically, a gauge zero mode couples to the conserved current of charged tower fields ($A^{(0)}_\mu J^\mu_{(n)}$), while the four-dimensional graviton couples to their energy--momentum tensor ($h^{(0)}_{\mu\nu} T^{\mu\nu}_{(n)}$). 

It is therefore important to determine whether the cutoff-independent $1/m_n$ scaling extends to these universal interactions. If the amplitude carries $k$ powers of the tower three-momentum, with $k=1$ for a current and $k=2$ for the transverse-traceless graviton, the radial part of the hard integral of Eq.~\eqref{eq:qint} becomes
\begin{equation}
\int_0^{q_0}\frac{q^{2+2k}\,\dd q}{(q^2+m_n^2)^2}=
\begin{cases}
\pi/4m_n, & k=0,\\
q_0-3\pi m_n/4, & k=1,\\
q_0^3/3-2m_n^2q_0+5\pi m_n^3/4, & k=2,
\end{cases}
\label{eq:derivcount}
\end{equation}
up to terms that vanish as $q_0\to\infty$. The integral converges and decreases with the mass of the level only for non-derivative couplings. Upon the introduction of derivative couplings, it is dominated by momenta near the cutoff, as in the quartic theory. 
By this power counting, the mutual information of the graviton and gauge zero modes with the tower is controlled by the ultraviolet completion. A complete calculation, including the sum over polarisations, is required to find the coefficients.

\paragraph{Ultraviolet completion.} The cubic and quartic results respond differently to an ultraviolet completion. The tail of the cubic theory is controlled by the hard integral of Eq.~\eqref{eq:qint}, which is dominated by momenta $|\mathbf q|\sim m_n$. For levels well below the string scale $M_s$ this region lies within the field-theory regime, so a string completion modifies $A_3$ and the $1/n$ dependence only through corrections suppressed by powers of $m_n/M_s$, and the logarithmic growth of the level sum persists up to $N\simeq M_sR$. The quartic $I_n$ is instead dominated by momenta near the cutoff, which in a string completion is the string scale itself; its power-law growth describes how it increases as the cutoff is raised within field theory, while its actual value is set by string-scale dynamics that the field-theory vertex does not capture. Above $M_s$ the Kaluza--Klein tower is joined by winding modes and string oscillator states, whose density grows exponentially with mass \cite{Hagedorn:1965st}, while string interactions at high energy are exponentially soft \cite{Gross:1987kza,Gross:1987ar}. Whether the contributions of these states to the mutual information converge depends on the competition between the two effects and is not settled by a field-theory calculation. For position-space entanglement, string theory has been argued to regulate the ultraviolet divergences of the entropy \cite{Balasubramanian:2018axm,Dabholkar:2023ows,Dabholkar:2024neq}, but those divergences arise from short distances near the entangling surface rather than from hard momenta in the environment, and the momentum-space partition considered here has, to our knowledge, not been studied in string theory.

\paragraph{Laboratory analogues.}\label{sec:lab} Effective dimensional reduction is not restricted to compact extra dimensions. In cold atomic gases confined by strongly anisotropic traps, motion along the tightly confined directions is quantised into transverse levels separated by a gap. Below the gap the gas is described by an effective theory in one or two dimensions, and the excited transverse levels play the role of the Kaluza--Klein tower~\cite{Olshanii:1998,Mazets:2008}. Interactions couple the low-energy sector to these levels and generically leave it in a mixed state. Information measures and their level dependence need not exactly follow the results obtained here, since the dispersion relation and the dimension of the low-energy sector differ from those of a relativistic field on a circle. The entanglement generated by dimensional reduction can nevertheless in principle be measured in such systems. The purity of a subsystem has been measured in cold atoms by interfering two identical copies \cite{Daley:2012xhf,Islam:2015}, although for spatial subsystems rather than for a partition by transverse level.

\paragraph{Open problems.} The calculation is carried out at leading order in the coupling, in flat space, for scalar fields on a circle, and several extensions follow directly. An immediate issue that follows from discussions in this paper is a complete gauge-invariant calculation for the graviton zero mode, and for gauge zero modes. The calculation can also be extended to de Sitter space. For $1/R\gg H$ the entanglement computed here is a static property of the adiabatic vacuum, so a decoherence rate for the curvature perturbation requires the regime $1/R\lesssim H$, in which the expansion excites the light levels and the calculation must be redone in de Sitter space. Repeating the calculation on $T^2$, $S^2$ or a general internal manifold, with the appropriate degeneracies and overlap integrals, would test whether the level sum grows as a power of $N$, as Weyl counting indicates. This is the relevant case for string compactifications, whose internal manifolds have dimension six or seven, and a treatment of the momentum-space partition in string theory itself would determine whether the contributions above the string scale are finite. The corresponding calculation for cold atoms in anisotropic traps, with the spectrum and selection rules of the trap, would show how the entanglement of the low-energy sector depends on the transverse level in a system where it can be measured (Section~\ref{sec:lab}). 
\section{Conclusions}
\label{sec:conclusions}

Kaluza--Klein reduction is justified by decoupling, which guarantees that the tower affects low-energy correlation functions only through local operators suppressed by its mass. Decoupling, however, is a statement about the low-energy observables of the four-dimensional theory, not about its state. We have studied the state directly, computing at leading order the entanglement entropy of the accessible zero-mode sector and its mutual information with a single Kaluza--Klein level for a five-dimensional scalar compactified on a circle, with a cubic or quartic self-interaction.

For free fields the vacuum is a product state across levels, so the reduced state of the zero modes is pure and any entanglement with the tower is generated by interactions. When the accessible sector couples linearly and without derivatives to pairs of tower modes, the contribution of this coupling to the mutual information with a level falls as $1/m_n$.
Whether this tail survives in the full expression for mutual information depends on the vertex. A cubic vertex with one leg in the accessible sector places its other two legs at levels $\pm n$. 
Consequently, for such a vertex, only one hard momentum is free, and the coefficient $A_3=(\mu R)^2/1024\pi^3$ is independent of the cutoff. Since each state involves a single level, the mutual informations with different levels add up to the mutual information with the whole tower, which grows as $\ln N$ with the number of levels. A quartic vertex produces the same states with a second soft quantum, which have the same tail, but its fourth leg can also carry a hard zero-mode quantum or a quantum at a different level. These states leave a second hard momentum free, so that the mutual information with a level is dominated by momenta near the cutoff and the $1/m_n$ tail is masked.

The corrections the tower makes to low-energy amplitudes are dominated by the lightest levels and are captured by local operators, hence the observables do not depend on how far the tower extends. However, the reduced state itself does not decouple this way, since every octave of levels contributes equally to the mutual information. This dependence is invisible to the effective theory. The entanglement resides in the time-nonlocal terms of the effective action that a derivative expansion discards, and it is measured by the entropy and the mutual information, which are nonlinear functionals of the reduced density matrix. The local four-dimensional effective theory therefore describes the low-energy observables of the reduced system but not its state. Unbroken supersymmetry, which cancels the vacuum energy of the compactification, does not remove this incompleteness, since at leading order it cannot cancel a sum of squared amplitudes; a supersymmetric Minkowski compactification still leaves the four-dimensional observer with a mixed state.

Several questions remain open. The couplings of the graviton and gauge zero modes, present in realistic compactifications, carry derivatives, and power counting places their mutual information with the tower near the cutoff. A complete calculation, including the polarisation sums, is needed to fix the coefficients. On internal manifolds of higher dimension, as in string compactifications, Weyl's law suggests that the level sum grows as a power of $N$ rather than logarithmically, which would make the effect parametrically larger. In de Sitter space, a tower heavier than the Hubble rate leaves the entanglement computed here as a static property of the adiabatic vacuum, while a lighter tower may contribute to the decoherence of the curvature perturbation, which requires the de Sitter calculation itself. In cold atomic gases in anisotropic traps the transverse levels play the role of the tower, so the entanglement generated by dimensional reduction can in principle be measured there. Finally, our results are valid at leading order in the coupling, which requires, $\varepsilon$ of Eq.~\eqref{eq:epsilon}, to be small. Determining what happens when $\varepsilon$ becomes of order unity requires going beyond this order. We regard this work as a first step towards understanding how much of the higher-dimensional state survives in a four-dimensional description, and which features of the compactification determine it.

\appendix

\section{Eigenvalues of the Reduced Density Matrix}
\label{app:eigen}

This appendix provides a self-contained derivation of the spectrum of the reduced density matrix $\rho_A$ quoted in Section~\ref{sec:bmv}.

\subsection*{Block Structure of the Reduced Density Matrix}

To evaluate the quantum correlations across the Hilbert space partition $\mathcal{H} = \mathcal{H}_A \otimes \mathcal{H}_B$, we first organise the unnormalised, first-order interacting vacuum state $|\tilde\Omega\rangle$ according to its multiparticle content:
\begin{equation}
|\tilde\Omega\rangle
= |0_A,0_B\rangle
+ \sum_{Y \neq 0} C_{Y0}\,|Y,0_B\rangle
+ \sum_{Z \neq 0} C_{0Z}\,|0_A,Z\rangle
+ \sum_{Y,Z \neq 0} C_{YZ}\,|Y,Z\rangle,
\qquad C_{X} = \mathcal O(\lambda),
\end{equation}
where $Y$ labels free multiparticle states in the accessible low-energy sector $\mathcal{H}_A$ (soft zero-mode quanta), and $Z$ labels free multiparticle states in the unobserved environment $\mathcal{H}_B$ (heavy tower modes and zero-mode quanta above the resolution scale). The amplitudes $C_{X} = -\langle X|H_I|0\rangle / E_X$ are determined by first-order perturbation theory in the coupling $\lambda$.

Tracing out the environment degrees of freedom in $\mathcal{H}_B$ using the orthonormality of the basis states ($\langle Z|Z'\rangle = \delta_{ZZ'}$) yields the unnormalised reduced density matrix $\tilde\rho_A = \Tr_B |\tilde\Omega\rangle\langle\tilde\Omega|$ on $\mathcal{H}_A$. In the basis $\{|0_A\rangle, |Y\rangle\}$, its matrix elements read:
\begin{align}
\langle 0 | \tilde\rho_A | 0 \rangle &= 1 + \sum_{Z \neq 0} |C_{0Z}|^2, \nonumber \\
\langle Y | \tilde\rho_A | 0 \rangle &= C_{Y0} + \sum_{Z \neq 0} C_{YZ} C^\ast_{0Z} = C_{Y0} + \mathcal O(\lambda^2), \nonumber \\
\langle Y | \tilde\rho_A | Y' \rangle &= C_{Y0} C^\ast_{Y'0} + \sum_{Z \neq 0} C_{YZ} C^\ast_{Y'Z} \equiv C_{Y0} C^\ast_{Y'0} + G_{YY'}.
\end{align}
In the form of a $2 \times 2$ block matrix separating the vacuum state $|0_A\rangle$ from the subspace of soft excited states $\{|Y\rangle\}$:
\begin{equation}
\tilde\rho_A = \begin{pmatrix} a & v^\dagger \\ v & D \end{pmatrix},
\qquad
a = 1 + \mathcal O(\lambda^2),
\quad
v_Y = C_{Y0} + \mathcal O(\lambda^2),
\quad
D = v v^\dagger\Big|_{\mathcal O(\lambda^2)} + G.
\end{equation}

The block $G_{YY'} \equiv \sum_{Z \neq 0} C_{YZ} C^\ast_{Y'Z} = \mathcal O(\lambda^2)$ is the \emph{Gram matrix} of the entangled amplitude vectors. Defining $w_Y \equiv (C_{YZ})_{Z \neq 0}$ as an amplitude vector residing in $\mathcal{H}_B$, each Gram matrix entry corresponds to an inner product of two such vectors, $G_{YY'} = \langle w_{Y'}, w_Y \rangle$. Because any Gram matrix is Hermitian ($G_{YY'}^* = G_{Y'Y}$) and positive-semidefinite ($\sum_{Y,Y'} x_Y^* G_{YY'} x_{Y'} = \|\sum_Y x_Y w_Y\|^2 \ge 0$), all of its eigenvalues are non-negative. This is the origin of the positivity argument of Section~\ref{sec:positivity}: no choice of symmetry or superpartner completion can render the eigenvalues of $G$ negative.

\subsection*{Schur Complement and Cancellation of Pure-$A$ Amplitudes}

We want to determine the eigenvalue spectrum of a block matrix $\begin{pmatrix} a & v^\dagger \\ v & D \end{pmatrix}$ where $a \approx 1$ is large compared to $v\sim \mathcal{O}(\lambda), D \sim \mathcal{O}(\lambda^2)$. The small eigenvalues of the full matrix are determined by the Schur complement of \(a\):
\begin{equation}
S_c = D - \frac{v \, v^\dagger}{a}.
\end{equation}
Substituting the explicit perturbative expressions into $S_c$ yields:
\begin{equation}
S_c = \Big( C_{Y0} C^\ast_{Y'0} + G_{YY'} \Big) - \frac{C_{Y0} C^\ast_{Y'0}}{1 + \mathcal{O}(\lambda^2)}
= \Big( C_{Y0} C^\ast_{Y'0} + G_{YY'} \Big) - C_{Y0} C^\ast_{Y'0}\Big[ 1 + \mathcal O(\lambda^2) \Big]
= G_{YY'} + \mathcal O(\lambda^4).
\end{equation}

The pure-$A$ amplitudes $C_{Y0}$ cancel out and small eigenvalues of the reduced density matrix are determined, up to order \(\lambda^2\), entirely by the Gram matrix \(G\).

The amplitudes \(C_{Y0}\) describe a pure state inside \(\mathcal{H}_A\). It is only a local deformation of the observable sector. Such a deformation can be removed by a unitary transformation acting only on \(A\). Since this cannot create entanglement, pure-\(A\) excitations do not contribute to the von Neumann entropy of \(\rho_A\).
Similarly, pure-\(B\) excitations \(C_{0Z}\) live entirely in the traced-out sector. They affect the normalisation but do not by themselves create entanglement visible in \(\rho_A\).

Only the mixed amplitudes \(C_{YZ}\), with excitations in both sectors, generate entanglement. This is precisely the information retained by the Gram matrix.

\subsection*{Spectrum, $g_Y$, and Basis Invariance of the Entropy}

We label the Gram matrix as $G = \lambda^2 \hat G$ so that $\hat{G} = \mathcal{O}(\lambda^0)$. Let $\{\gamma_i\}$ denote the non-negative eigenvalues of $\hat G$. Since $\rho_A$ has unit trace, the spectrum of the normalised reduced density matrix $\rho_A$ to order $\lambda^2$ consists of:
\begin{equation}
p_0 = 1 - \lambda^2 \sum_i \gamma_i,
\qquad
p_i = \lambda^2 \gamma_i \quad (\gamma_i \ge 0).
\label{eq:app-spectrum}
\end{equation}

The quantity $g_Y$ used in Section~\ref{sec:bmv} represents the diagonal entry of $\hat G$ evaluated in the free multiparticle basis,
\begin{equation}
g_Y \equiv \hat G_{YY} = \frac{1}{\lambda^2} \sum_{Z \neq 0} |C_{YZ}|^2.
\end{equation}

We now evaluate the von Neumann entropy $S_{\rm vN} = -p_0 \ln p_0 - \sum_i p_i \ln p_i$. Expanding 
\[
-p_0 \ln p_0 = \lambda^2 \sum_i \gamma_i + \mathcal{O}(\lambda^4),
\qquad
-p_i \ln p_i
=
-\lambda^2 \gamma_i
\ln(\lambda^2 \gamma_i).
\]
and substituting into $S_{\rm vN}$,
\begin{align}
S_{\rm vN} 
&= \lambda^2 \sum_i \gamma_i - \sum_i \lambda^2 \gamma_i \left( \ln\lambda^2 + \ln\gamma_i \right) + \mathcal{O}(\lambda^4 \ln\lambda) \nonumber \\
&= \lambda^2 \left[ \ln\frac{1}{\lambda^2} + 1 \right] \sum_i \gamma_i + \lambda^2 \sum_i \gamma_i \ln\frac{1}{\gamma_i} + \mathcal{O}(\lambda^4 \ln\lambda).
\label{eq:Sappendix}
\end{align}

The coefficient of the leading 
logarithm $\ln(1/\lambda^2)$ is the sum of the eigenvalues,
\begin{equation}
\sum_i \gamma_i = \Tr \hat G = \sum_Y \hat G_{YY} = \sum_Y g_Y = \frac{1}{\lambda^2} \sum_{Y,Z \neq 0} |C_{YZ}|^2 \equiv \Sigma.
\end{equation}

Because the matrix trace is strictly invariant under any change of basis ($\Tr(U \hat{G} U^\dagger) = \Tr \hat{G}$), the leading-logarithm coefficient $\Sigma$ does not depend on the off-diagonal elements of $\hat{G}$, nor does it require diagonalising the infinite-dimensional matrix $\hat{G}$. This trace invariance allows the calculations in Sections~\ref{sec:22}--\ref{sec:phi3} to reduce directly to standard phase-space integrals (Eqs.~\eqref{eq:I22}--\eqref{eq:Imix}) without solving a complex matrix eigenvalue problem.

\subsection*{Purity deficit at order $\lambda^2$}
\label{app:purity}

The purity deficit is simpler than the von Neumann entropy because its leading term is fixed entirely by the trace of the Gram matrix. Using the spectrum in Eq.~\eqref{eq:app-spectrum},
\begin{equation}
\Tr\rho_A^2=p_0^2+\sum_i p_i^2
=\left(1-\lambda^2\sum_i\gamma_i\right)^2
+\lambda^4\sum_i\gamma_i^2+\cdots
=1-2\lambda^2\Sigma
+\lambda^4\left[\Sigma^2+\sum_i\gamma_i^2\right]+\cdots ,
\end{equation}
where the ellipsis denotes contributions not fixed by the order-$\lambda^2$ spectrum. Since $\gamma_i\ge0$,
\begin{equation}
\sum_i\gamma_i^2\le\left(\sum_i\gamma_i\right)^2=\Sigma^2,
\end{equation}
so the displayed $\lambda^4$ correction is bounded by $2\lambda^4\Sigma^2$. The leading order purity deficit is therefore
\begin{equation}
1-\Tr\rho_A^2=2\lambda^2\Sigma+\mathcal O(\lambda^4),
\qquad
S_2=-\ln\Tr\rho_A^2=2\lambda^2\Sigma+\mathcal O(\lambda^4).
\end{equation}

Unlike Eq.~\eqref{eq:Sappendix}, this result contains no $\ln(1/\lambda^2)$ and depends on the spectrum of $\hat G$ only through $\Sigma=\Tr\hat G$, so it is unaffected by how the continuum of small eigenvalues is discretised (Section~\ref{sec:twoentropies}).

\section{Entanglement Kernel}
\label{app:norm}

In this Appendix, we provide the complete field-theoretic derivation of the entanglement kernel $\mathcal{I}$.  

\subsection{Quantisation and Spatial Wave-Function Integration}
\label{app:box_integration}

To maintain rigorous control over phase-space volume factors, we quantise the 4D field components $\phi_n(x)$ in a spatial box of volume $V$ with periodic boundary conditions. The 4D field operators are expanded in discrete plane-wave modes as
\begin{equation}
    \phi_n(\mathbf{x}, t) = \sum_{\mathbf{k}} \frac{1}{\sqrt{2\omega_{\mathbf{k},n} V}} \left( a_{\mathbf{k},n} e^{i(\mathbf{k} \cdot \mathbf{x} - \omega_{\mathbf{k},n} t)} + a_{\mathbf{k},-n}^\dagger e^{-i(\mathbf{k} \cdot \mathbf{x} - \omega_{\mathbf{k},n} t)} \right),
    \label{eq:app_box_field_expansion}
\end{equation}
The relativistic dispersion relation is $\omega_{\mathbf{k},n} = \sqrt{\mathbf{k}^2 + m_n^2}$, with $m_n = |n|/R$. Creation and annihilation operators satisfy canonical commutation relations:
\begin{equation}
    [a_{\mathbf{k},n}, a_{\mathbf{k}',n'}^\dagger] = \delta_{\mathbf{k}, \mathbf{k}'} \, \delta_{n, n'}.
\end{equation}

Inserting the mode expansion into the 5D action and integrating over $y$ yields the 4D interaction Hamiltonian (Section~\ref{sec:setup}):
\begin{equation}
    H_I(t) = \frac{\lambda}{4!} \int_V d^3x \sum_{n_1+n_2+n_3+n_4=0} \phi_{n_1}(\mathbf{x},t) \phi_{n_2}(\mathbf{x},t) \phi_{n_3}(\mathbf{x},t) \phi_{n_4}(\mathbf{x},t),
\end{equation}
where $\lambda \equiv \lambda_4 = \lambda_5 / (2\pi R)$ is the dimensionless 4D coupling constant.

\subsection{Transition matrix}
\label{app:matrix_element}

Consider a four-particle state $|X\rangle = |(\mathbf{k}_1, n_1), (\mathbf{k}_2, n_2), (\mathbf{k}_3, n_3), (\mathbf{k}_4, n_4)\rangle$, 
the transition matrix element between the free vacuum and this state is
\begin{equation}
    \langle X | H_I | 0 \rangle = \frac{\lambda}{4!} \int_V d^3x \sum_{\sum n_i' = 0} \langle 0 | a_{\mathbf{k}_4, n_4} a_{\mathbf{k}_3, n_3} a_{\mathbf{k}_2, n_2} a_{\mathbf{k}_1, n_1} \, \phi_{n_1'}(\mathbf{x}) \phi_{n_2'}(\mathbf{x}) \phi_{n_3'}(\mathbf{x}) \phi_{n_4'}(\mathbf{x}) | 0 \rangle.
\end{equation}
 For four distinct quanta, there are $4! = 24$ fully equivalent ways to contract the creation operators $a_{\mathbf{k}_i, n_i}^\dagger$ with the field expansion. These 24 terms  cancel the $1/4!$ prefactor in $H_I$. The spatial integral of the four plane waves gives $\int_V d^3x\, e^{i\sum_i\mathbf k_i\cdot\mathbf x}=V\delta_{\sum\mathbf k_i,\mathbf 0}$, so the transition matrix element is
\begin{equation}
    \langle X | H_I | 0 \rangle = \lambda V\delta_{\sum n_i, 0}  \, \delta_{\sum \mathbf{k}_i, \mathbf{0}} \, \prod_{i=1}^4 \frac{1}{\sqrt{2\omega_i V}}.
    \label{eq:app_matrix_element}
\end{equation}

Squaring Eq.~\eqref{eq:app_matrix_element} yields
\begin{equation}
    |\langle X | H_I | 0 \rangle|^2 = \frac{\lambda^2}{16 \omega_1 \omega_2 \omega_3 \omega_4 V^2} \, \delta_{\sum \mathbf{k}_i, \mathbf{0}} \, \delta_{\sum n_i, 0}.
    \label{eq:app_matrix_elem_squared}
\end{equation}

\subsection{Form of the Entanglement Kernel}
\label{app:sum_states_volume}

The quantity controlling the leading perturbative entanglement entropy is the dimensionless trace
\[
\Sigma
=
\sum_{X\in AB}
\frac{|\langle X|H_I|0\rangle|^2}{\lambda^2 E_X^2},
\]
where \(E_X\) is the total free energy of the intermediate state \(|X\rangle\). The sum is over states containing excitations in both the accessible low-energy sector \(A\) and the unobserved environment \(B\). Only such mixed-sector states generate entanglement entropy.
From Eq.~\eqref{eq:app_matrix_elem_squared}, summing over ordered labels before the symmetry factors of Appendix~\ref{app:symmetry_factors} are applied,
\begin{equation}
    \Sigma = \sum_{X \in AB} \frac{|\langle X | H_I | 0 \rangle|^2}{\lambda^2 E_X^2} = \sum_{\{n_i\}}\delta_{\sum n_i, 0}\sum_{\mathbf{k}_1, \mathbf{k}_2, \mathbf{k}_3, \mathbf{k}_4} \left( \frac{\delta_{\sum \mathbf{k}_i, \mathbf{0}}}{16 \omega_1 \omega_2 \omega_3 \omega_4 V^2 E_X^2} \right),
\end{equation}
where $E_X = \sum_{i=1}^4 \omega_i$ is the total energy of the intermediate state.

Using the momentum-conserving delta to eliminate one momentum leaves three independent momentum sums. In the continuum limit,
\[
\sum_{\mathbf{k}_1}
\sum_{\mathbf{k}_2}
\sum_{\mathbf{k}_3}
\longrightarrow
\frac{V^3}{(2\pi)^9}
\int d^3k_1\,d^3k_2\,d^3k_3 .
\]

Combining the factor \(V^3\) from the three independent momentum sums with the factor \(V^{-2}\) from the squared matrix element gives
\[
\Sigma
=
V
\sum_{\{n_i\}}\delta_{\sum n_i, 0}\int
\frac{d^3k_1\,d^3k_2\,d^3k_3}{(2\pi)^9}
\frac{1}{16\omega_1\omega_2\omega_3\omega_4 E_X^2}.
\]
Thus \(\Sigma\) is proportional to the spatial volume.
Dividing by $V$ yields the intensive entanglement kernel $\mathcal{I} \equiv \Sigma / V$:
\begin{equation}
    \mathcal{I} = \frac{\Sigma}{V} = \sum_{\{n_i\}}\delta_{\sum n_i, 0} \int \frac{d^3k_1 d^3k_2 d^3k_3}{(2\pi)^9} \frac{1}{16 \omega_1 \omega_2 \omega_3 \omega_4 E_X^2} .
    \label{eq:app_kernel_intensive}
\end{equation}

\subsection{Channel Combinatorics and Symmetry Factors}
\label{app:symmetry_factors}
KK number conservation ($\sum n_i= 0$) restricts the entangling processes that involve the tower to the three channels below; the zero-mode-only configurations are those of four-dimensional $\lambda\phi^4$ and are discussed in Section~\ref{sec:channels}.

\subsubsection{The $(2,2)$ Channel ($\phi_0 \phi_0 \phi_n \phi_{-n}$)}
This channel entangles two soft zero-mode quanta in subsystem $\mathcal{H}_A$ ($|\mathbf{k}_{1,2}| \le \mu$) with a heavy KK pair in subsystem $\mathcal{H}_B$. 
 Exchange of the two identical soft zero-mode quanta $(\mathbf{k}_1 \leftrightarrow \mathbf{k}_2)$ describes the same physical state in $\mathcal{H}_A$, introducing a combinatorial symmetry factor of $1/2!$.
    The heavy tower quanta carry distinct opposite KK charges $+n$ and $-n$. Restricting the KK level sum to $n \ge 1$ ensures that each charged pair $(+n, -n)$ is counted exactly once when the tower momentum $\mathbf{q}_1$ is integrated freely over $\mathbb{R}^3$.
Applying symmetry factor $1/2!$ yields the explicit kernel $\mathcal{I}_{22}$:
\begin{equation}
    \mathcal{I}_{22} = \frac{1}{2} \sum_{n \ge 1} \int_{|\mathbf{k}_1|, |\mathbf{k}_2| \le \mu} \frac{d^3k_1 d^3k_2 d^3q_1}{(2\pi)^9} \frac{1}{16 \omega_1 \omega_2 \Omega_1 \Omega_2 E^2},
    \label{eq:app_I22_final}
\end{equation}
where $\omega_{1,2} = |\mathbf{k}_{1,2}|$, $\Omega_{1,2} = \sqrt{\mathbf{q}_{1,2}^2 + n^2/R^2}$, momentum conservation fixes $\mathbf{q}_2 = -(\mathbf{k}_1 + \mathbf{k}_2 + \mathbf{q}_1)$, and $E = \omega_1 + \omega_2 + \Omega_1 + \Omega_2$.

\subsubsection{The $(1,3)$ Channel ($\phi_0 \phi_{n_1} \phi_{n_2} \phi_{n_3}$)}
This channel entangles one soft zero-mode quantum in $\mathcal{H}_A$ ($|\mathbf{k}| \le \mu$) with three heavy tower quanta in $\mathcal{H}_B$ ($n_i \neq 0, \sum n_i = 0$).
 Permuting the three identical field operators in the tower triplet yields $3! = 6$ indistinguishable state assignments, introducing a symmetry prefactor of $1/3!$.
 With this, the kernel $\mathcal{I}_{13}$ takes the form
\begin{equation}
    \mathcal{I}_{13} = \frac{1}{3!} \sum_{\substack{n_1+n_2+n_3=0 \\ n_i \neq 0}} \int_{|\mathbf{k}| \le \mu} \frac{d^3k \, d^3q_1 d^3q_2}{(2\pi)^9} \frac{1}{16 \omega \Omega_1 \Omega_2 \Omega_3 E^2},
    \label{eq:app_I13_final}
\end{equation}
where $\omega = |\mathbf{k}|$, $\Omega_i = \sqrt{\mathbf{q}_i^2 + n_i^2/R^2}$, momentum conservation fixes $\mathbf{q}_3 = -(\mathbf{k} + \mathbf{q}_1 + \mathbf{q}_2)$, and $E = \omega + \Omega_1 + \Omega_2 + \Omega_3$.

\subsubsection{The mixed channel ($\phi_0^{\rm soft}\phi_0^{\rm hard}\phi_n\phi_{-n}$)}
This channel entangles one soft zero-mode quantum in $\mathcal H_A$ ($|\mathbf k|\le\mu$) with a hard zero-mode quantum ($|\mathbf k_0|>\mu$) and a KK pair in $\mathcal H_B$. The two zero-mode quanta lie in disjoint momentum regions, so exchanging them does not produce a second assignment within the integration domain and no symmetry factor arises; restricting to $n\ge1$ counts each tower pair once, as in the $(2,2)$ channel. Hence
\begin{equation}
\mathcal I_{\rm mix}=\sum_{n\ge1}\int_{|\mathbf k|\le\mu}\int_{|\mathbf k_0|>\mu}
\frac{\dd^3k\,\dd^3k_0\,\dd^3q_1}{(2\pi)^9}\,
\frac{1}{16\,\omega\,\omega_0\,\Omega_1\Omega_2\,E^2},
\label{eq:app_Imix_final}
\end{equation}
with $\omega=|\mathbf k|$, $\omega_0=|\mathbf k_0|$, $\mathbf q_2=-(\mathbf k+\mathbf k_0+\mathbf q_1)$ and $E=\omega+\omega_0+\Omega_1+\Omega_2$.

\section{Numerical evaluation}
\label{app:mc}

This appendix describes how the kernels of Sections~\ref{sec:22}--\ref{sec:phi3} were evaluated and how the constants quoted in the text were obtained. Units $R=1$ are used, so that $m_n=n$, and $\mu R=1$ unless stated otherwise; dimensionful results follow by restoring $R$.

\paragraph{Method.} The kernels are phase-space integrals over the momenta of the intermediate state, weighted by the squared amplitude and the energy denominator, with the remaining momenta fixed by three-momentum conservation. They were evaluated by importance-sampled Monte Carlo integration. Soft zero-mode momenta were drawn uniformly from the ball $|\mathbf k|\le\mu$. Unbounded hard momenta were drawn isotropically from the radial density $p(q)=(2/\pi)\,s/(s^2+q^2)$, with $s$ set to the typical hard scale of the configuration. At large $q$ this density falls as $q^{-2}$, like the radial integrand $q^2f(q)\propto q^{-2}$, so that the ratio of integrand to density, whose average gives the estimate, stays bounded and the statistical error decreases as $N_{\rm s}^{-1/2}$ with the number of samples $N_{\rm s}$. In the quartic theory at finite cutoff, the five-dimensional cutoff of Section~\ref{sec:inclusive} is imposed on every tower leg and $|\mathbf k_0|\le\Lambda_{\rm UV}$ on the hard zero mode. Statistical errors are below one per cent for the cubic and the combined $(2,2)$ and mixed channels, and below two per cent for the $(1,3)$ channel.

\begin{figure}[t]
\centering
\includegraphics[width=\textwidth]{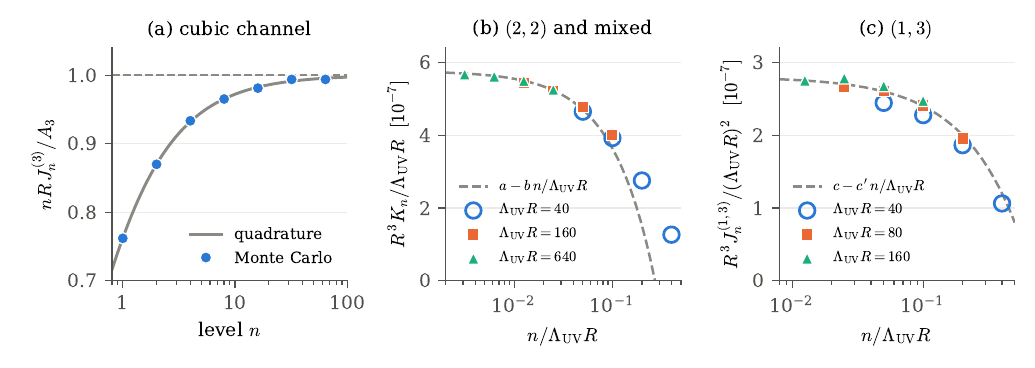}
\caption{Numerical evaluation of the kernels, for $\mu R=1$. (a) Cubic channel: $nJ_n^{(3)}$ in units of $A_3/R$, from Monte Carlo (points) and direct quadrature (line). Dashed line corresponds to the asymptotic value. (b) Combined $(2,2)$ and mixed channels: $K_n/\Lambda_{\rm UV}R$ for three different cutoffs, compared with the linear form of Eq.~\eqref{eq:Knlin} (dashed), valid for $n\ll\Lambda_{\rm UV}R$. (c) $(1,3)$ channel: $J_n^{(1,3)}/(\Lambda_{\rm UV}R)^2$ for three different cutoffs, compared with Eq.~\eqref{eq:J13lin} (dashed). The statistical errors are below $1\%$ in (a) and (b) and below $2\%$ in (c).}
\label{fig:mc}
\end{figure}

\paragraph{Cubic channel.} The kernel has one soft and one free hard momentum. Figure~\ref{fig:mc}(a) shows $nJ_n^{(3)}$ in units of its asymptotic value $A_3/R$, Eq.~\eqref{eq:A3}. The Monte Carlo values agree with a direct quadrature of the same integral to within $0.3\%$ at all levels, and both approach $A_3$ from below. The correction, of relative order $\mu R/n$, comes from the soft momentum in the energy denominator.

\paragraph{$(2,2)$ channel.} The computed $J_n$ approach $A/nR^3$, Eq.~\eqref{eq:Acoeff}, from below; a power-law fit to levels $13$--$20$ gives $J_n\propto n^{-0.97\pm0.02}$, and runs at $\mu R=0.5$ and $2$ reproduce the $(\mu R)^4$ scaling.

\paragraph{$(2,2)$ and mixed channels combined.} Equation~\eqref{eq:Kn} was evaluated at $\Lambda_{\rm UV}R=40$, $160$ and $640$. Plotted as $K_n/\Lambda_{\rm UV}R$ against $n/\Lambda_{\rm UV}R$, Fig.~\ref{fig:mc}(b), the results for different cutoffs fall on a single curve, which approaches the linear form of Eq.~\eqref{eq:Knlin} for $n\lesssim0.05\,\Lambda_{\rm UV}R$. The intercept agrees with the value of $a$ obtained by quadrature of $F(0)$, and the slope approaches the value of $b$ obtained from Eq.~\eqref{eq:beta} as the cutoff is increased. Runs at $\mu R=0.5$ and $1$ confirm the $(\mu R)^2$ scaling.

\paragraph{$(1,3)$ channel.} The spectator level was sampled uniformly over the values allowed by the cutoff, and the two free hard momenta uniformly from the ball $|\mathbf q|\le\Lambda_{\rm UV}$. Plotted as $J_n^{(1,3)}/(\Lambda_{\rm UV}R)^2$ against $n/\Lambda_{\rm UV}R$, Fig.~\ref{fig:mc}(c), the results for $\Lambda_{\rm UV}R=40$, $80$ and $160$ follow the linear form of Eq.~\eqref{eq:J13lin}, with an intercept $c$ that increases slowly with the cutoff and approaches a limit as the cutoff is raised, and a slope $c'$ nearly independent of it.

\paragraph{Constants.} The constants used in the text are collected in Table~\ref{tab:constants}.

\begin{table}[H]
\centering
\small
\begin{tabular}{lllp{7.2cm}}
\toprule
constant & equation & value & obtained from \\
\midrule
$A_3$ & \eqref{eq:A3} & $3.15\times10^{-5}$ & analytic; reproduced by Monte Carlo and quadrature at large $n$ \\
$a$ & \eqref{eq:Knlin} & $5.76\times10^{-7}$ & quadrature of $F(0)$; Monte Carlo at $\Lambda_{\rm UV}R=640$ gives $5.7\times10^{-7}$ \\
$b$ & \eqref{eq:Knlin} & $2.1\times10^{-6}$ & quadrature of $\Phi$, Eq.~\eqref{eq:beta}; Monte Carlo slope is $1.9\times10^{-6}$ at $\Lambda_{\rm UV}R=640$ \\
$c$ & \eqref{eq:J13lin} & $\approx3\times10^{-7}$ & Monte Carlo only: $2.6$, $2.75$, $2.8\times10^{-7}$ at $\Lambda_{\rm UV}R=40$, $80$, $160$ \\
$c'$ & \eqref{eq:J13lin} & $\approx4\times10^{-7}$ & Monte Carlo only: $4.0$--$4.5\times10^{-7}$ at the same cutoffs \\
\bottomrule
\end{tabular}
\caption{Coefficients of the asymptotic forms used in the text, for $\mu R=1$. The Monte Carlo values at finite cutoff approach the analytical result as the cutoff is raised. $A_3$, $a$ and $b$ are known to about $1\%$ error, $c$ and $c'$ to about $10\%$.}
\label{tab:constants}
\end{table}

\paragraph{Data availability statement.} All numerical results of this paper, including the values in table~\ref{tab:constants} and figure~\ref{fig:mc}, were obtained with a Python script that is publicly available at \url{https://github.com/udaythattarampilly/kk-momentum-space-entanglement}. No other data were generated or analysed in this study.

\acknowledgments
The authors acknowledge the use of artificial-intelligence tools in this work. Anthropic's Claude, and Google AI Studio were used to check symbolic derivations, to write and cross-check the code for the Monte Carlo evaluations, and to draft and edit the manuscript. The authors take full responsibility for the scientific content of the paper.

\bibliographystyle{JHEP}
\bibliography{Ref1}

\end{document}